\makeatletter
\def\cl@chapter{}
\makeatother

\RequirePackage{fix-cm}
\documentclass[twocolumn,epjc3]{svjour3}

\usepackage[utf8x]{inputenc}
\usepackage[english]{babel}
\usepackage[intlimits]{amsmath}
\usepackage{amssymb}
\usepackage[artemisia]{textgreek}
\usepackage{upgreek}
\usepackage[abbreviations=true,retain-explicit-plus]{siunitx}
\usepackage{hyperref}
\usepackage[noabbrev]{cleveref}
\usepackage{xspace}
\usepackage{graphicx}
\usepackage{caption}
\usepackage{subcaption}
\usepackage{booktabs}

\newlength{\figurewidthonecol}
\newlength{\figurewidthtwocols}
\newcommand{\che}[1]{}

\newcommand{\mnu}{ {m_{\upnu}} }
\newcommand{\mnusq}{ {m_{\upnu}^2} }
\newcommand{\hmnu}{ \widehat{m_{\upnu}} }
\newcommand{\hmnusq}{ \widehat{m_{\upnu}^2} }
\newcommand{\me}{ {m_\text{e}} }
\newcommand{\upe}{\text{e}}

\newcommand{\va}{\ensuremath{V\!-\!A}\xspace}
\newcommand{\eps}{\ensuremath{\mathcal{E}}}
\newcommand{\Tcal}{\ensuremath{\mathcal{T}}}
\newcommand{\Ncal}{\ensuremath{\mathcal{N}}}

\newcommand{\tb}{\textbeta\;\ignorespaces}
\newcommand{\tbd}{\textbeta-\ignorespaces}

\newcommand*\chem[1]{\ensuremath{\mathrm{#1}}}

\newcommand{\kB}{k_\text{B}}

\newcommand{\vM}{v_\text{M}}

\newcommand{\Ecms}{E_\text{cms}}
\newcommand{\Elab}{E_\text{lab}}
\newcommand{\gcms}{\gamma_\text{cms}}

\newcommand{\vecms}{v_{\upe,\text{cms}}}
\newcommand{\velab}{v_{\upe,\text{lab}}}

\newcommand{\plong}{p_\parallel}
\newcommand{\ptrans}{p_\perp}
\newcommand{\Elong}{E_\parallel}
\newcommand{\Etrans}{E_\perp}

\newcommand{\tritPur}{\epsilon_\chem{T}}

\newcommand{\+}{\ensuremath{\;\;}}

\newcommand{\td}{\,\mathrm{d}}	

\DeclareSIUnit\percentcl{\percent\,C.L.}
\DeclareSIUnit\sigmacl{$\sigma$\,C.L.}
\DeclareSIUnit[number-unit-product = \,]{\permille}{\textperthousand}
\DeclareSIUnit\cps{cps}
\DeclareSIUnit\mcps{\milli\cps}

\journalname{Eur. Phys. J. C}

\begin{document}

\title{\tbd Decay Spectrum, Response Function and Statistical Model for Neutrino Mass Measurements with the KATRIN Experiment}
\titlerunning{Model for Neutrino Mass Measurements with the KATRIN Experiment}

\institute{%
     {Institute of Experimental Particle Physics (ETP), Karlsruhe Institute of Technology, Karlsruhe, Germany}\label{kit-iekp}
\and {Institute for Nuclear Physics (IKP), Karlsruhe Institute of Technology, Karlsruhe, Germany}\label{kit-ikp}
\and {Nuclear Science Division, Lawrence Berkeley National Laboratory, Berkeley, CA, USA}\label{lbnl}
\and {Laboratory for Nuclear Science and Department of Physics, Massachusetts Institute of Technology, Cambridge MA, USA}\label{mit}
\and {Max-Planck-Institut f\"{u}r Physik, F\"{o}hringer Ring 6, 80805 M\"{u}nchen, Germany}\label{mpi}
\and {Technische Universit\"{a}t M\"{u}nchen, James-Franck-Str. 1, 85748 Garching, Germany}\label{tum}
\and {Institut für Kernphysik, Westfälische Wilhelms-Universität Münster, Münster, Germany}\label{wwu}
}

\author{%
     {M.~Kleesiek}\thanksref{email1,fn1,kit-iekp}
\and {J.~Behrens}\thanksref{email2,kit-ikp}
\and {G.~Drexlin}\thanksref{kit-iekp}
\and {K.~Eitel}\thanksref{kit-ikp}
\and {M.~Erhard}\thanksref{kit-iekp}
\and {J.~A.~Formaggio}\thanksref{mit}
\and {F.~Glück}\thanksref{kit-ikp}
\and {S.~Groh}\thanksref{kit-iekp}
\and {M.~Hötzel}\thanksref{kit-iekp}
\and {S.~Mertens}\thanksref{mpi,tum}
\and {A.~W.~P.~Poon}\thanksref{lbnl}
\and {C.~Weinheimer}\thanksref{wwu}
\and {K.~Valerius}\thanksref{kit-ikp}
}

\thankstext{email1}{\email{marco.kleesiek@kit.edu}}
\thankstext{email2}{\email{jan.behrens@kit.edu}}
\thankstext{fn1}{n\'{e} M.~Haag}

\date{Received: date / Accepted: date}

\maketitle

\begin{abstract}
The objective of the Karlsruhe Tritium Neutrino (KATRIN) experiment is to determine the effective electron neutrino mass $m(\upnu_\upe)$ with an unprecedented sensitivity of \SI{0.2}{\eV/c^2} (\SI{90}{\percentcl}) by precision electron spectroscopy close to the endpoint of the \tbd decay of tritium.
We present a consistent theoretical description of the \tbd electron energy spectrum in the endpoint region, an accurate model of the apparatus response function, and the statistical approaches suited to interpret and analyze tritium \tbd decay data observed with KATRIN with the envisaged precision.
In addition to providing detailed analytical expressions for all formulae used in the presented model framework with the necessary detail of derivation, we discuss and quantify the impact of theoretical and experimental corrections on the measured $m(\upnu_\upe)$.
Finally, we outline the statistical methods for parameter inference and the construction of confidence intervals that are appropriate for a neutrino mass measurement with KATRIN. In this context, we briefly discuss the choice of the \tbd energy analysis interval and the distribution of measuring time within that range.
\end{abstract}

\setcounter{secnumdepth}{2}
\setcounter{tocdepth}{2}
\tableofcontents

\section{Introduction}
\label{sec:intro}

While neutrino oscillation experiments~\cite{bib:superk,bib:sno,bib:kamland} have provided unambiguous evidence of non-zero neutrino masses, the absolute neutrino mass scale remains an open question.
The primary objective of the Karlsruhe Tritium Neutrino (KATRIN) experiment is to probe this scale in a direct kinematic measurement at an unprecedented sensitivity of \SI{0.2}{\eV/c^2} (\SI{90}{\percentcl})~\cite{KATRIN2005}. The measurement principle is based on a shape analysis of the tritium \tbd decay spectrum by high precision electron spectroscopy. A non-zero neutrino mass will cause a distortion in the observed spectrum, which is most pronounced close to the endpoint energy of \SI{18.6}{keV}.
This technique has been successfully established by the direct neutrino mass experiments in Mainz and Troitsk, which place the most stringent direct upper limit on the effective electron neutrino mass~\cite{Kraus2005,Aseev2011,Aseev2012,Olive2014}:
\begin{equation}
    m(\upnu_\upe) < \SI{2}{eV/c^2} \quad (\SI{95}{\percentcl}) \; .
\end{equation}
Improving this limit in $m(\upnu_\upe)$ by a factor of 10 demands an enhancement in statistical and systematic precision of the effective observable $m^2(\upnu_\upe)$ by a factor of 100. This requires both an in-depth understanding of the theoretical electron \tbd decay spectrum and an accurate knowledge of the experimental response in measuring the spectral shape. In \cref{sec:setup} we explain the KATRIN setup in more detail.

It is the goal of this work to provide a complete and up-to-date model of the experiment, such that it can be used as either a prescription or reference for upcoming analyses of tritium \tbd decay data observed with KATRIN. For established aspects of this model, we refer to the appropriate publications. For those not yet published at all or not in the required detail, we provide the necessary derivations. The later will mostly be the case for the description of the experimental response function, which has been considerably refined during recent commissioning phases.

In this work we first present a detailed account of the theoretical \tb spectrum of tritium, with an emphasis on molecular effects in \chem{T_2} (\cref{sec:diff}). We then outline the experimental configuration of KATRIN (\cref{sec:setup}), before we elaborate on the individual characteristics that define the response of our instrument in \cref{sec:response}. The statistical techniques suited to determine the effective neutrino mass from a fit of the modeled \tb spectrum to the measured data are treated in \cref{sec:measurement}. A summary of this work is given in \cref{sec:conclusion}.

Throughout this article we use natural units ($c = \hbar = 1$) for better readability, except for \cref{sec:doppler,sec:cyclotron} where we use SI units instead.

\section{Theoretical description of the differential \tbd decay spectrum}
\label{sec:diff}

In this section we compile a comprehensive analytical description of the differential \tbd decay spectrum, with specific focus on gaseous molecular tritium \chem{T_2}, the \tb emitter used by KATRIN. We will also evaluate the relevance of various theoretical correction terms on the neutrino mass analysis.

In the following, we use the shorthand notation $\mnu = m(\upnu_\upe)$ for better readability. Furthermore, we assume there is no difference between the masses of the neutrinos and the anti-neutrinos, i.e.\ $\mnu = m(\upnu_\upe) = m(\bar{\upnu}_\upe)$.

In the \tbd decay of atomic tritium, the surplus energy $Q$ is shared between the electron's kinetic energy $E$, the total neutrino energy and the recoil energy $E_\text{rec}$ of the much heavier daughter nucleus:
\begin{equation}
    \chem{T} \, \longrightarrow \, \chem{^3He}^+ + \text{e}^- + \bar{\upnu}_\text{e} \, + \, Q(\chem{T}) \; .
\end{equation}
In the case of a vanishing neutrino mass, the electron spectrum would terminate at the endpoint energy
\begin{equation}
    E_0 = Q - E_\text{rec} \; .
\end{equation}

\subsection{Fermi theory}

The differential decay rate of a tritium nucleus can be described with Fermi's Golden Rule as~\cite{Otten2008}
\begin{align}
        \nonumber
    \left( \frac{\td \Gamma}{\td E} \right)_\text{nuc} &= \frac{G_\text{F}^2 \, |V_\text{ud}|^2}{2\pi^3} \; |M_\text{nuc}|^2 \; F(Z, E) \; p\, (E+\me) \\
      &\quad \cdot \sum_i \; |U_{\upe i}|^2 \; \epsilon \; \sqrt{ \epsilon^2 - m_i^2} \; \Theta(\epsilon - m_i) \; .
    \label{eq:diff}
\end{align}
The Fermi coupling constant $G_\text{F}$ is projected onto the (u,\,d) coupling by the Cabibbo angle $\theta_\text{C}$ with $|V_\text{ud}| = \cos{\theta_\text{C}} = \num{0.97425(22)}$~\cite{Olive2014}.

For tritium \tbd decay -- a super-allowed transition -- the nuclear transition matrix element $M_\text{nuc}$ is independent of the electron energy. It can be divided into a vector (Fermi) part and an axial (Gamow-Teller) part
\begin{equation}
    \left|M_\text{nuc}\right|^2 = g_\text{V}^2 + 3 g_\text{A}^2 \; ,
\end{equation}
with the vector coupling constant $g_\text{V} = 1$ and the axial-vector coupling constant defined by $g_\text{A}/g_\text{V} = \num{-1.2646(35)}$ in tritium~\cite{bib:akulov}.

The classical Fermi function $F(Z, E)$ accounts for the Coulomb interaction between the outgoing electron and the daughter nucleus with atomic charge $Z$ (here $Z = 2$):
\begin{equation}
    \label{eq:fermi}
    F(Z, E) = \frac{2\pi\eta}{1-\exp(-2\pi\eta)}
\end{equation}
with the Sommerfeld parameter $\eta = \alpha Z/\beta$; $\alpha$ is the fine structure constant and $\beta = v/c$ is the electron velocity relative to speed of light. Here $F(Z,E)$ is written in the non-relativistic approximation; the relativistic $F(Z,E)_\text{rel}$ and its commonly-used approximation is given in \cref{sec:appendix:fermi}.

The full spectrum is an incoherent sum over the three known neutrino mass eigenstates $m_i$ ($i=1,2,3$) with the intensity of each component defined by the squared magnitude of the neutrino mixing matrix elements $|U_{\upe i}|^2$~\cite{Robertson1988}.

The phase-space factor of the outgoing electron with momentum $p$ is given by the factor $p\, (E+\me)$. The phase space of the emitted neutrino is the product of the neutrino energy $\epsilon = E_0 - E$ and the neutrino momentum $\sqrt{ \epsilon^2 - m_i^2}$, which determines the shape of the \tbd electron spectrum near the tritium endpoint $E_0$. The Heaviside step function $\Theta$ ensures that the kinetic energy cannot become negative.

\begin{figure}[ht]
    \centering
    \includegraphics[width=\figurewidthonecol]{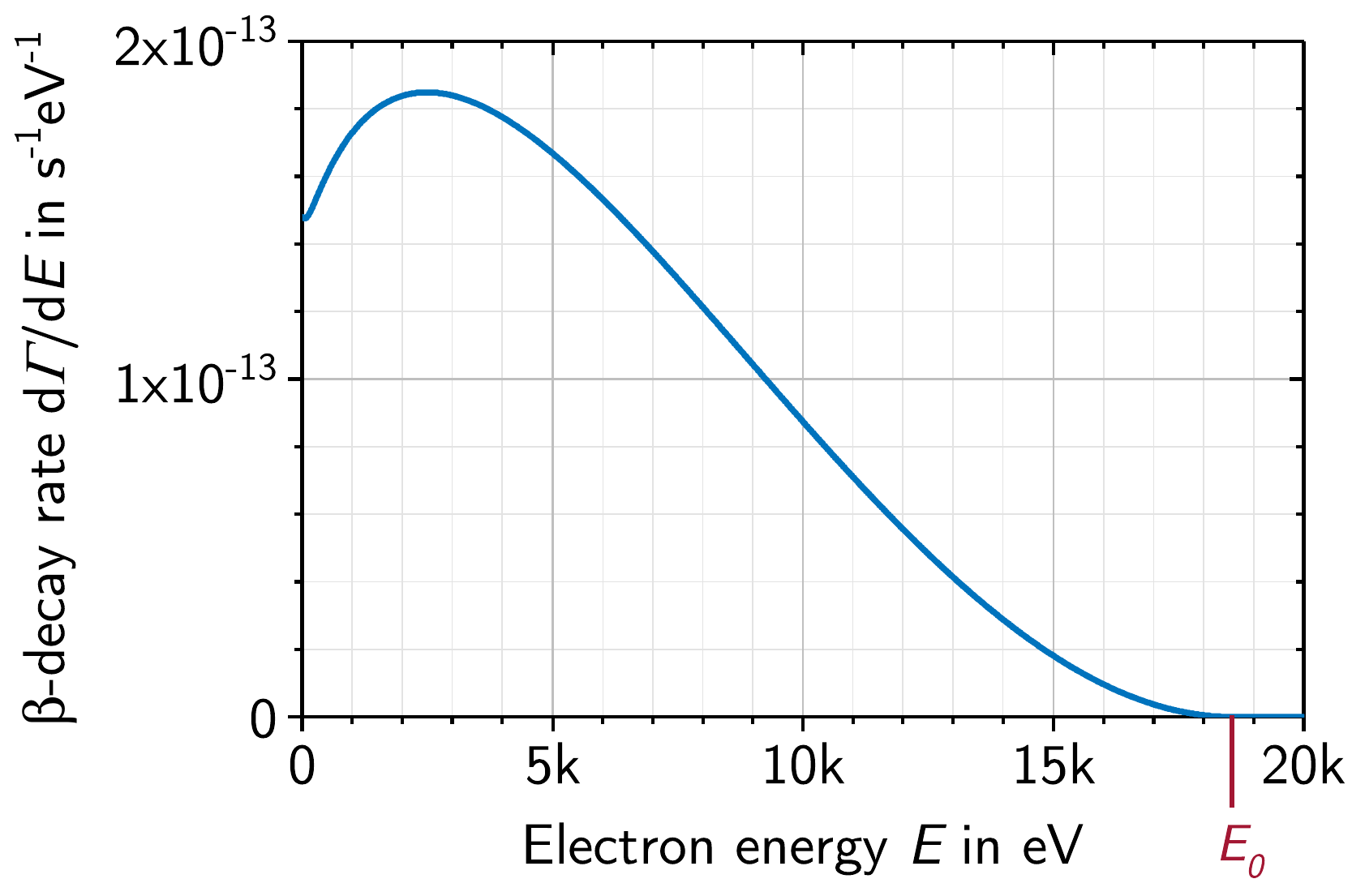}
    \caption[Differential \tbd decay spectrum]{%
    The differential \tbd electron energy spectrum for the \tbd decay of molecular tritium with the endpoint energy $E_0$ of \SI{18.574}{keV}.
    The given units correspond to the decay rate of a single tritium nucleus.
    }
    \label{fig:diff_complete}
\end{figure}

\begin{figure}[ht]
    \centering
    \includegraphics[width=\figurewidthonecol]{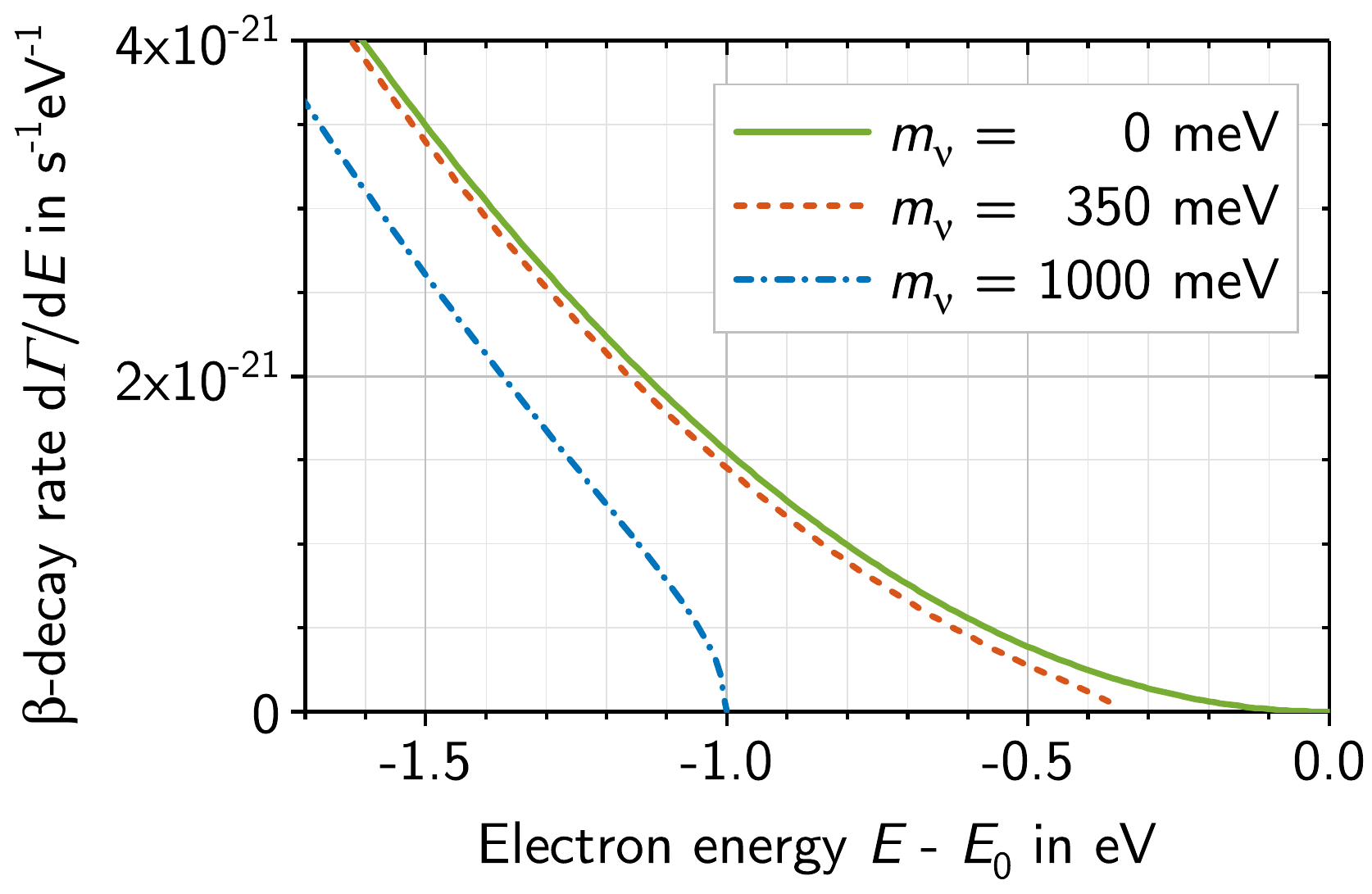}
    \caption[Differential \tbd decay spectrum at the endpoint]{%
    The differential \tbd electron energy spectrum near the endpoint for the decay of molecular tritium as given by \cref{eq:diff}, under the assumption of various neutrino masses $\mnu$.}
    \label{fig:diff_endpoint}
\end{figure}

The full \tbd decay spectrum is shown in \cref{fig:diff_complete}. The dependence of the spectral shape on the effective neutrino mass close to the endpoint is depicted in \cref{fig:diff_endpoint}.

\subsection{Neutrino mass eigenstate splittings}

In the KATRIN sensitivity range we can simplify the analysis by considering the effective electron neutrino mass square $\mnusq$ of a quasi-degenerate model in \cref{eq:diff}, given by an incoherent sum as
\begin{equation}
    \mnusq = \sum_{i} |U_{\upe i}|^2 \, m_i^2 \; .
\end{equation}
Calculations have shown this approximation of the \tbd decay spectrum to be valid, both for the normal and inverted mass hierarchies~\cite{bib:shrock,Farzan2003}.

\subsection{Molecular tritium \chem{T_2}}

When we consider the \tbd decay of gaseous molecular tritium \chem{T_2},
\begin{equation}
    \chem{T_2} \, \longrightarrow \, \chem{^3HeT}^+ + \text{e}^- + \bar{\upnu}_\text{e} \, + \, Q(\chem{T_2}) \; ,
\end{equation}
the released energy $Q$ has to be corrected for the differences in electronic binding energies between the atomic and actual molecular systems (see~\cite{Otten2008} for a detailed explanation). The nuclear recoil also excites a spectrum of rotational and vibrational final states in the daughter molecular system, and generates excitations of its electronic shell. The neutrino energy in \cref{eq:diff} has to be corrected by
\begin{equation}
    \epsilon \; \rightarrow \; \epsilon_f = E_0 - V_f - E \; ,
\end{equation}
with the endpoint $E_0(\chem{T_2}) = \SI{18574.00(7)}{eV}$ for molecular tritium~\cite{Otten2008,Myers2015}. The recoil energy reaches a maximum of $E_\text{rec} = \SI{1.72}{eV}$ at the \tbd endpoint, which gives a fixed endpoint energy $E_0(\chem{T_2}) = Q(\chem{T_2}) - E_\text{rec}$~\cite{Otten2008}.
The differential decay rate, with the additional summation over each final state $f$ with energy $V_f$ and weighing by the transitional probability $P_f$ to a state $f$ in the daughter molecule, is then:
\begin{align}
        \nonumber
    \frac{\td \Gamma}{\td E} &= \frac{G_\text{F}^2 \, |V_\text{ud}|^2}{2\pi^3} \; |M_\text{nuc}|^2 \; F(Z, E) \cdot p\, (E+\me) \\
      &\quad \cdot \sum_f \; P_f \; \epsilon_f \; \sqrt{ \epsilon_f^2 - \mnusq } \; \Theta(\epsilon_f - \mnu) \; .
    \label{eq:diff2}
\end{align}

\subsection{Excited molecular final states}
\label{sec:fsd}

\begin{figure*}[tb]
    \centering
    \includegraphics[width=\figurewidthtwocols]{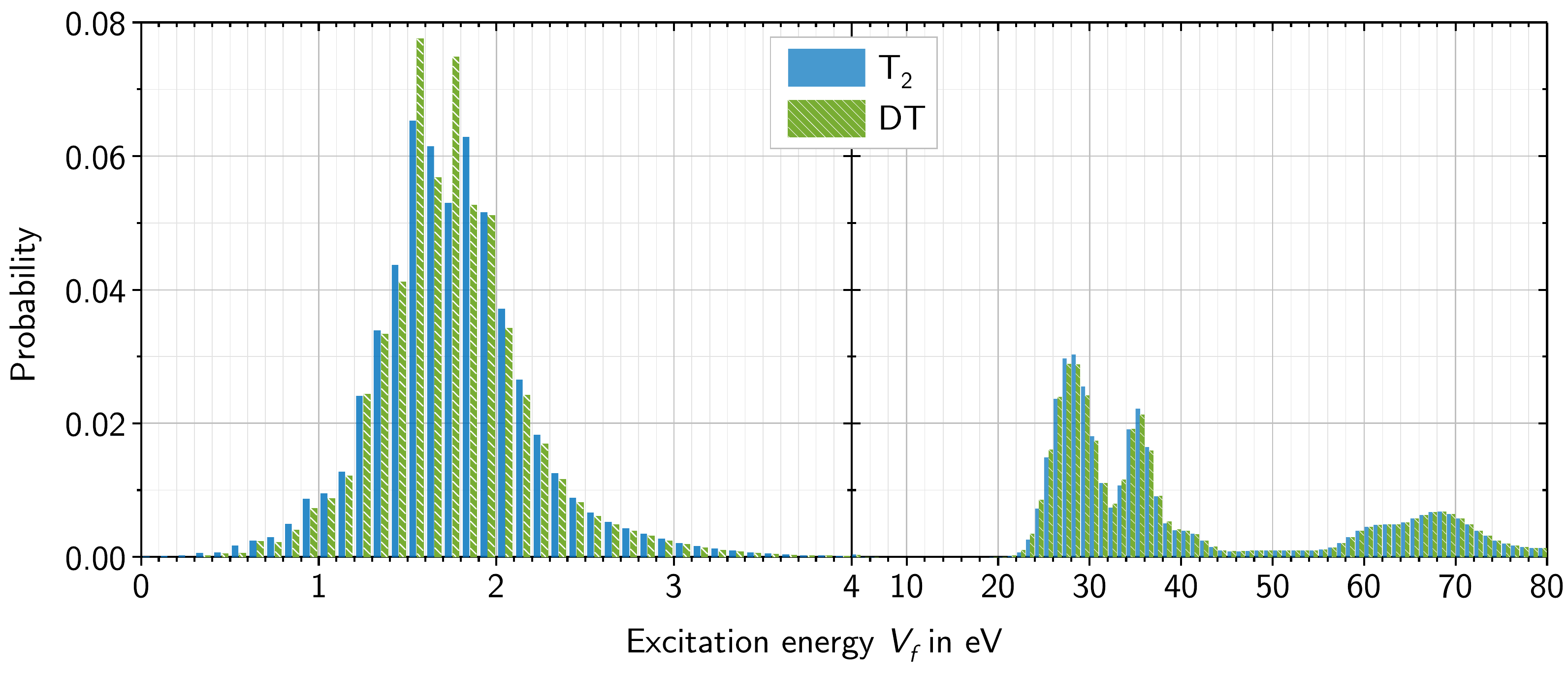}
    \caption[Final-State Distribution]{%
        Comparison of molecular final-state distributions of \chem{HeT^+} and \chem{HeD^+}. Sampled from~\cite{Doss2006,Doss2008} with a \SI{0.1}{eV} binning for excitation energies $V_f \leq \SI{4}{eV}$ and a \SI{1.0}{eV} binning for $V_f > \SI{4}{eV}$, summed over the initial angular momenta states $0 \leq J \leq 2$ according to their population at a temperature of $T = \SI{30}{K}$.
    }
    \label{fig:fsd}
\end{figure*}

After the decay, the daughter molecular system is left in an excited rotational, vibrational and electronic state. According to theoretical calculations, about \SI{57}{\percent} of all \chem{T_2} \tbd decays result in the rovibronically-broadened electronic ground state with an average excitation energy of about \SI{1.7}{eV}, while the others go to the excited electronic states~\cite{Jonsell1999}. Each discrete final state effectively branches into its own \tb spectrum with a distinct endpoint energy.

The accuracy of a neutrino mass measurement critically depends on the knowledge of the distribution of these final states, which have to be taken from theory. Precise calculations of the final state distributions of the hydrogen isotopologues (\chem{T_2 \rightarrow HeT^+}, \chem{DT \rightarrow HeD^+} and \chem{HT \rightarrow HeH^+}) have been performed in the endpoint region~\cite{Doss2006,Doss2008}.
The discrete energy states and their transition probabilities have been determined below the dissociation threshold, while continuous distributions are available above the threshold. A comprehensive review of the theory of the tritium final-state spectrum and current validation efforts can be found in~\cite{Bodine2015}.

\Cref{fig:fsd} gives a comparison of the final-state distributions of \chem{HeT^+} and \chem{HeD^+}. The differences in their distributions arise from the mass difference; thus, a precise knowledge of the source gas isotopological composition and its stabilization on the \SI{0.1}{\percent} level are necessary. Laser Raman spectroscopy~\cite{Fischer2012} provides two important input parameters for our source model: the tritium purity $\tritPur$ denoting the fraction of tritium nuclei\footnote{If we denote the fraction of all hydrogen isotopologues $X$ by $c(X)$ with $\sum_X c(X) = 1$, then the tritium purity is given by $\tritPur = c(\chem{T_2}) + c(\chem{DT}) / 2 + c(\chem{HT}) / 2$.}, and $\kappa$ denoting the ratio of \chem{DT} versus \chem{HT}.

In the calculations provided by~\cite{Doss2006,Doss2008,Saenz2000}, the higher recoil energies of the lighter isotopologues are incorporated into their respective energy spectra that are given relative to the recoil energy of \chem{HeT^+}. That way, the final-state distributions of each isotopologue can be summed and weighted according to its abundance in the source gas.
Furthermore, these calculations provide separate distributions for each initial quantum state of molecular angular momentum, denoted by the quantum number $J$.
These must be weighted according to the population of their respective $J$ states before the \tbd decay, which is given by a Boltzmann distribution
\begin{equation}
\label{eq:angular}
    P_J(T) \propto g_s g_J \exp\left( - \frac{\Delta E_J}{k_B T}\right) \; ,
\end{equation}
where $T$ is the local temperature of the source gas, $k_B$ the Boltzmann constant and $\Delta E_J$ the energy to the electronic ground state. The rotational degeneracy of the distribution is given by the factor $g_J = (2J+1)$, whereas $g_s$ accounts for the spin degeneracy of the nuclei. It is $g_s = 1$ for heteronuclear molecules (\chem{DT}, \chem{HT}) without spin coupling. For \chem{T_2} as a homonuclear molecule, it is given by the ratio $\lambda$ of molecules in an ortho (parallel nuclei spins) state or the ratio $1\!-\!\lambda$ in the para states (anti-parallel nuclei spins). Hence, $g_s = \lambda$ for ortho states with odd $J$ and $g_s = 1\!-\!\lambda$ for para states with even $J$~\cite{bib:souers}.
In the KATRIN tritium circulation system the source gas is forced into thermal equilibrium at $T = \SI{300}{K}$ by a permeator membrane%
    \footnote{The gas is then injected into the source beam tube and rapidly cooled down to \SI{30}{K}. Because the gas spends only a short time ($\lesssim \SI{1.5}{s}$) at this temperature, the rotational states cannot equilibrate again.}%
, resulting in $\lambda \simeq 0.75$~\cite{Bodine2015}.

\subsection{Exact relativistic three-body calculation}

\begin{figure}[ht]
    \centering
    \includegraphics[width=\figurewidthonecol]{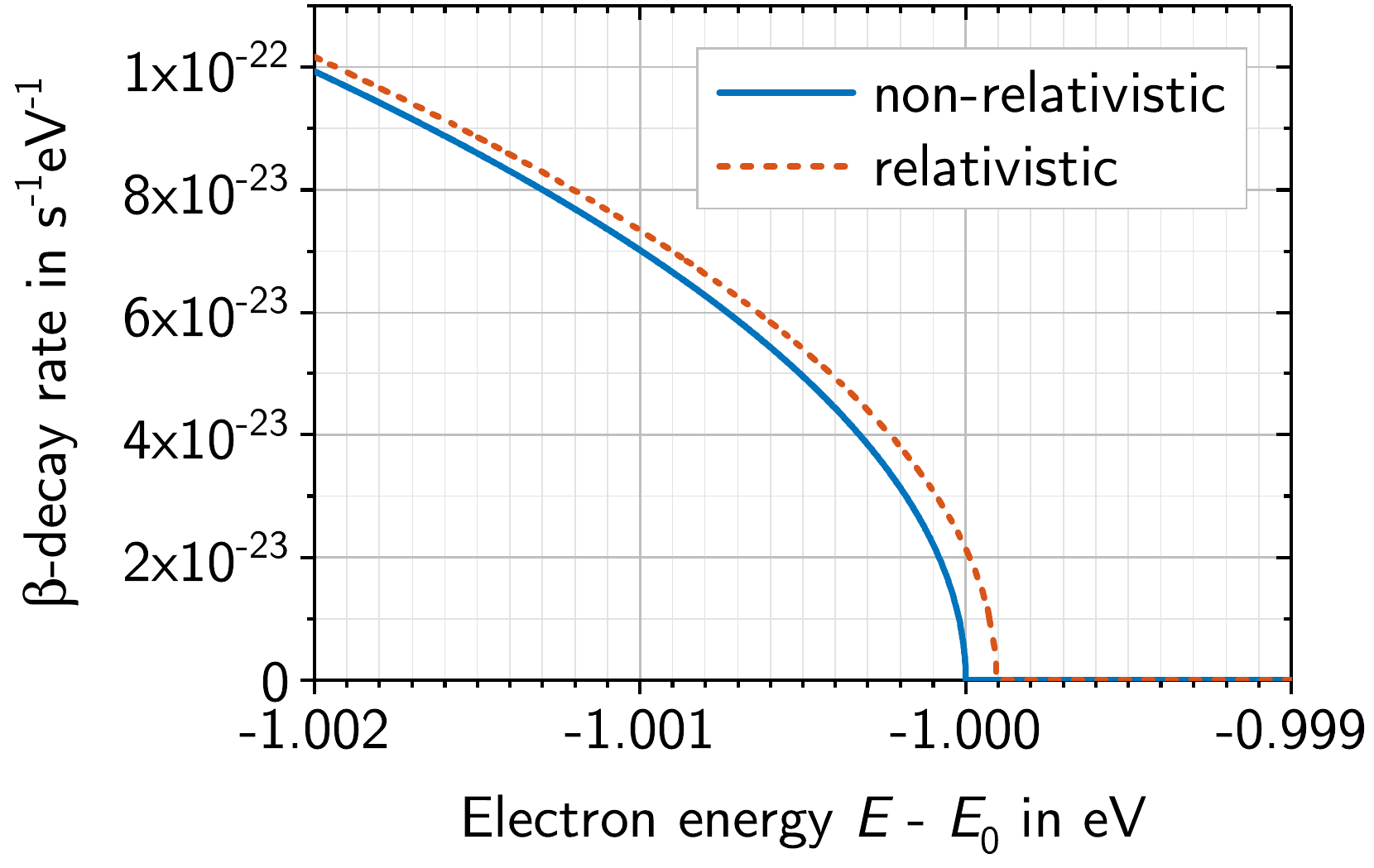}
    \caption[Non-Relativistic Approximation]{%
    Comparison of the differential \tbd electron energy spectrum of atomic tritium for the full relativistic kinematic treatment and the non-relativistic approximation, assuming a neutrino mass of $\mnu = \SI{1}{eV}$.
    }
    \label{fig:diff_relativistic}
\end{figure}

The \tb spectrum formalism outlined above contains approximations to the exact relativistic calculations of the three-body phase space density~\cite{Simkovic2008,Masood2007}. In deriving \cref{eq:diff2}, the dependence of the daughter molecule's recoil energy $E_\text{rec}$ on the neutrino mass $m_i$ and the final-state spectrum $V_f$ is neglected.
This approximation results in a minute shift of the maximum electron energy, which is on the order of \SI{0.1}{meV}~\cite{Masood2007}, as depicted in \cref{fig:diff_relativistic}. In the neutrino mass analysis, such a shift in the energy scale is compensated by the external constraint of the endpoint $E_0$; thus, the effective two-body representation of \cref{eq:diff2} is an adequate approximation in the energy region of interest (also see \cref{tab:systematics}). A summary of the energy-dependent, higher-order correction terms is given in \cref{sec:corrections}.

\subsection{Additional correction terms}
\label{sec:corrections}

In addition to the Fermi function ($F(Z, E)$) correction factors arising from other nuclear and atomic physics effects must be evaluated and applied multiplicatively. The formulae and the references to these effects are given in \cref{sec:appendix:corr}. The following is a synopsis.

\begin{itemize}
    \item Radiative corrections: In addition to the Coulomb interaction described by $F(Z, E)$, electromagnetic effects involving contributions from virtual and real photons give rise to a correction factor $G(E,E_0)$.

    \item Screening: The unscreened $F(Z, E)$, which describes the Coulomb interaction between the daughter nucleus and the departing \tbd electron, must be corrected by a factor  $S(Z,E)$ that accounts for the screening effect on the Coulomb field by the $\text{1s}$-orbital electrons left behind by the parent molecule.

    \item Recoil effects: In the relativistic elementary particle treatment of the \tbd decay (see for instance~\cite{Wu1983,Masood2007}), energy-dependent recoil effects on the order of $1/M$ can be calculated, with $M$ being the mass of \chem{^3He}. These effects --- spectrum shape modification due to a three-body phase space, weak magnetism and \va interference --- are typically combined into a common factor $R(E, E_0, M)$.

    \item Finite structure of the nucleus: Because the \chem{^3He^+} daughter nucleus is not a point-like object, the Coulomb field does not scale with an inverse-squared relationship within the radius, leading to a correction factor $L(Z,E)$. A proper convolution of the electron and neutrino wave functions with the nucleonic wave function throughout the nuclear volume leads to another factor $C(Z,E)$.

    \item Recoiling Coulomb field: The departing electron does not propagate in the field of a stationary charge, but one which is itself recoiling from the electron emission. This effect introduces another correction factor $Q(Z, E, E_0, M)$.

    \item Orbital-electron interactions: A correction factor $I(Z,E)$ is introduced to account for possible quantum mechanical interactions between the departing \tbd electron and the $\text{1s}$-orbital electrons.

\end{itemize}
The differential \tb spectrum, including all the theoretical correction factors discussed above, can be written as follows:
\begin{align}
        \nonumber
    \left(\frac{\td \Gamma}{\td E} \right)_\text{C} &= \frac{G_\text{F}^2 \, |V_\text{ud}|^2}{2\pi^3} \; (g_\text{V}^2 + 3 g_\text{A}^2) \; F_\text{rel}(Z, E) \\
        \nonumber
      &\quad \cdot p \, (E+\me) \cdot S \, L \, C \, I \\
      &\quad \cdot \sum_f \; G \, R \, Q \cdot P_f \, \epsilon_f \, \sqrt{\epsilon_f^2 - \mnusq} \; \Theta(\epsilon_f - \mnu) \; .
    \label{eq:diff3}
\end{align}
The corrections connected to the recoil of the daughter nucleus, namely $R$ and $Q$, and the radiative corrections $G$, depend on the endpoint energy and the phase space of a specific excited final state. This dependency is reflected in \cref{eq:diff3}, as these factors are summed over the possible final states.

\begin{figure*}[h!]
    \centering
    \includegraphics[width=\figurewidthtwocols]{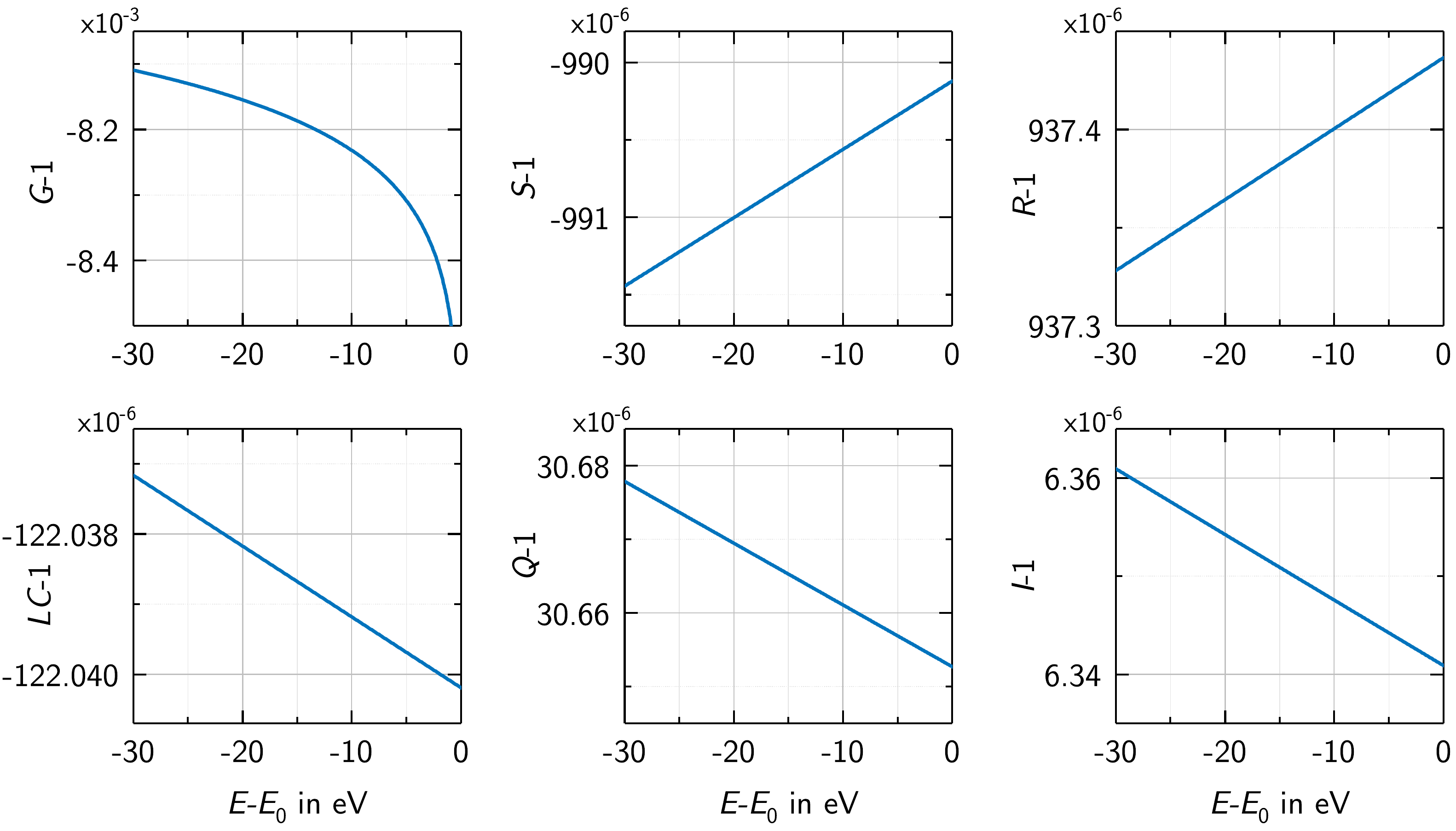}
    \caption[Theoretical Corrections]{%
    Theoretical correction factors to the differential \tbd decay spectrum of \chem{T_2}, evaluated in an interval $30\,\text{eV}$ below the endpoint $E_0$ and summed over possible final states.
    }
    \label{fig:corrections}
\end{figure*}

In \cref{fig:corrections}, a graphical overview of these correction factors in the energy interval \SI{30}{eV} below the tritium endpoint is given. The radiative corrections have the most significant effect with a pronounced energy dependence, as they deplete the spectrum completely towards the endpoint. Most other corrections are negligible in the neutrino mass analysis, as further detailed in \cref{sec:impact} and \cref{tab:systematics}.

\section{The KATRIN experiment}
\label{sec:setup}

The experimental setup of KATRIN combines a high-luminosity windowless gaseous molecular tritium source (WGTS) with an integrating electrostatic spectrometer of MAC-E filter (magnetic adiabatic collimation with electrostatic filter) type~\cite{Beamson1980,Lobashev1985,Picard1992}, offering a narrow filter width and a wide solid-angle acceptance at the same time.

\begin{figure*}[ht!]
    \centering
    \includegraphics[width=.9\linewidth]{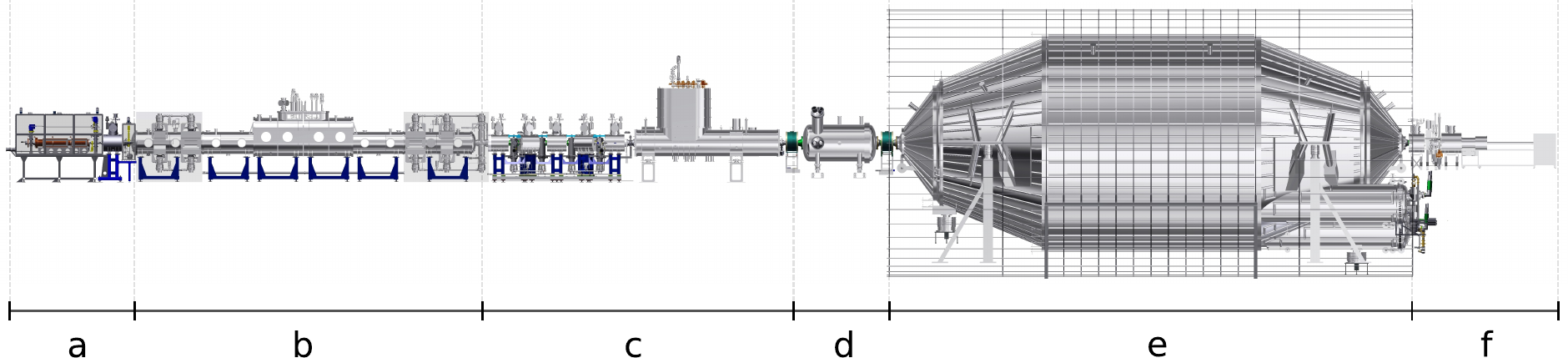}
    \caption[KATRIN experimental setup]{%
        The KATRIN experimental setup, \SI{70}{m} in length.
        The monitoring and calibration section (a) residing at the rear of the high-luminosity windowless source (b) provides stable and precise monitoring of tritium gas properties.
        The transport system (c) magnetically guides the electrons further downstream and prevents tritiated gas from entering the spectrometer section, which features two spectrometers operating as MAC-E-filters. The smaller pre-spectrometer (d) acts as a pre-filter for low energy electrons, and the larger main spectrometer (e) is used for the energy analysis in the endpoint region. A segmented detector (f) acts as a counter for the transmitted signal electrons.
    }
    \label{fig:katrin}
\end{figure*}

The apparatus depicted in \cref{fig:katrin} features several major subsystems.
The isotopological composition, temperature, and density fluctuations of the tritium source are monitored by a set of calibration devices housed in the rear section (a).
The windowless gaseous tritium source (b) contains a beam tube of length $L = \SI{10}{m}$ and diameter $d = \SI{90}{mm}$, residing in a nominal magnetic field of \SI{3.6}{T}, where re-purified molecular tritium (\chem{T_2}) is continuously circulated by injection at the center and pumping at both ends through a closed loop system~\cite{Sturm2010,Schloesser2013c,Priester2015}.
To prevent tritiated gas from entering the spectrometer section, the transport section (c) combines differential pumping with cryogenic pumping to reduce the tritium flow by 14~orders of magnitude~\cite{Gil2010,Lukic2012}.
The \tbd electrons are guided through the entire beamline by a magnetic field~\cite{arXivArenz2018e} into the pre-spectrometer (d), which acts as a pre-filter that blocks the low-energy electrons of the \tbd spectrum~\cite{Prall2012}. The energy analysis around the endpoint region takes place in the main spectrometer (e), which is operated under ultra-high vacuum conditions~\cite{Arenz2016} at a retarding voltage of about \SI{-18.6}{kV}. Both spectrometers are designed as MAC-E filters, and the main spectrometer achieves a very narrow filter width ($\lesssim \SI{1}{eV}$)~\cite{Otten2008} while providing high luminosity for the \tbd electrons.
Electrons with sufficient energy pass both the MAC-E filters and are then counted at a segmented silicon PIN diode detector (f)~\cite{Amsbaugh2015} with 148 individual pixels. An integrated \tbd spectrum is recorded by scanning the retarding voltage in the endpoint region.

\subsection{MAC-E filter principle}
\label{sec:mace}

The electrons emitted isotropically from tritium \tbd decay in the gaseous source are guided adiabatically by magnetic fields. In the forward direction the \tbd electrons are confined in cyclotron motion along the magnetic field lines towards the MAC-E filter.
Along their path to the \emph{analyzing plane} (central plane) of the spectrometer, the magnetic field strength decreases by several orders of magnitude%
    \footnote{The KATRIN main spectrometer employs a set of air coils to allow fine-shaping of the weak guiding field in the analyzing plane, and to compensate for influences by the earth's magnetic field and solenoid fringe fields~\cite{Glueck2013,Erhard2018}.}%
. Due to the conservation of magnetic moment in a slowly varying field, most of the electrons' transverse momentum is adiabatically transformed into longitudinal momentum. With a high negative potential ($U \approx \SI{-18.6}{kV}$, corresponding to the endpoint energy of tritium) at its center and most of the electron momentum being parallel to the magnetic field lines, the MAC-E filter acts as an electrostatic high-pass energy filter. Only electrons with positive longitudinal energy (the kinetic energy in direction of the magnetic field line) along their entire trajectory are transmitted, while the others are reflected and re-accelerated towards the entrance of the spectrometer.

The residual transverse energy, which cannot be analyzed by the filter, is defined by the ratio of the maximum $B_\text{max}$ to the minimum magnetic field $B_\mathrm{min} = B_\text{A}$. This key characteristic of the MAC-E filter is commonly called the \emph{filter width} (or sometimes energy resolution)
\begin{equation}
    \Delta E = \frac{B_\text{A}}{B_\text{max}} \cdot E \, \frac{\gamma\!+\!1}{2} \; ,
    \label{eq:FilterWidth}
\end{equation}
with $E$ being the electron kinetic energy and $\gamma = \frac{E}{m_\upe} + 1$ the relativistic gamma factor with the electron rest mass $m_\upe$.

\section{Response function of the KATRIN experiment}
\label{sec:response}

In the KATRIN experiment, the energy of the \tbd electrons is analyzed using the MAC-E filter technique as described in \cref{sec:setup}. For a specific electrostatic retardation potential $U$, the count rate of electrons at the detector can be calculated, given the probability of an electron with a starting energy $E$ to traverse the whole apparatus and hit the detector. This probability is described by the so-called transmission function $T(E,U)$. Additional modifications arise from energy loss and scattering in the source, and reflection of signal electrons propagating from their point of origin until detection. These effects are incorporated together with the transmission function into the response function $R(E,U)$, which is vital for the neutrino mass analysis as it describes the propagation of signal electrons that contribute to the integrated \tbd spectrum.

For illustrative purposes, we first consider a source containing a given number of tritium nuclei ($N_\chem{T}$) that decay with an isotropic angular distribution\footnote{%
    At a temperature of \SI{30}{K} and a magnetic field strength of \SI{3.6}{T}, the polarization of the tritium nuclei can be neglected.}%
. The emitted electrons are guided by magnetic fields through the spectrometer. The detection rate at the detector for a given spectrometer potential $U$ can be expressed as:
\begin{equation}
    \dot N(U) = \frac{1}{2}\,N_\chem{T}\int_{qU}^{E_0} \frac{\td \Gamma}{\td E}(E_0, m_\upnu^2) \cdot R(E, U) \, \td E \; ,
\end{equation}
where the factor of $\frac{1}{2}$ incorporates the fact that the response function $R(E,U)$ only considers electrons emitted in the forward direction.

In the following, an analytical description of the response function of the KATRIN experiment will be laid out. At first, we derive the transmission function of the MAC-E filter that is implemented by the main spectrometer (\cref{sec:transmission}). In \cref{sec:eloss} we consider energy loss in the source and develop a first description of the response function. Inhomogeneities in the MAC-E filter (\cref{sec:transmission_inhomogeneity}) and the source (\cref{sec:source}) requires extension of the model by a segmentation of the source and spectrometer volume. Further modifications to the response function arise from considering the effective source column density which an individual \tbd electron traverses (\cref{sec:scattering}), changes to the electron angular distribution (\cref{sec:response_noniso}), thermal motion of the source gas (\cref{sec:doppler}), and energy loss by cyclotron radiation (\cref{sec:cyclotron}).
After discussing these contributions, in \cref{sec:integrated_spectrum} we arrive at a description of the integrated spectrum that is measured by the KATRIN experiment. We close the discussion with a general note on experimental energy uncertainties (\cref{sec:escale}) and give a quantitative overview of theoretical corrections and systematic effects (\cref{sec:impact}) on the neutrino mass analysis.

\subsection{Transmission function of the MAC-E filter}
\label{sec:transmission}

The transmission of \tbd electrons through the MAC-E filter is an important characteristic of the measurement and a significant part of the response function. In the simplest case, one can assume that electrons enter the MAC-E filter with an isotropic angular distribution and propagate adiabatically towards the detector. In the discussion here we apply the adiabatic approximation (see \cref{eq:response:adiabatic_approx} below), which is fulfilled in the case of KATRIN.

In general, an electron from the source will reach the detector if the momentum $\plong$ parallel to the magnetic field lines (or the corresponding fraction $\Elong$ of the kinetic energy) is always positive. The transformation of transverse to parallel momentum and back in a slowly varying magnetic field $B$ is governed by the following adiabatic invariant (which corresponds to the conserved orbital momentum $\mu = \Etrans / B$ in the non-relativistic limit):
\begin{equation}
    \frac{p^2_{\perp}}{B} = \text{const.}
    \label{eq:response:adiabatic_approx}
\end{equation}
In the following discussion we use the general relation between the transverse momentum $\ptrans$ of an electron with its transverse kinetic energy $\Etrans$:
\begin{equation}
    \ptrans^2 = \Etrans \; (\gamma + 1) \cdot \me
\end{equation}
with the relativistic gamma factor $\gamma = \frac{E}{\me} + 1$, and thereby define the transverse kinetic energy as:
\begin{equation}
    \Etrans = E \; \sin^2 \theta \; .
\end{equation}
Similarly, we define the longitudinal kinetic energy as $\Elong = E \; \cos^2 \theta$. The polar angle $\theta = \angle(\vec{p},\vec{B})$ of an electron momentum to the magnetic field is called the \emph{pitch angle}.

We can now define the adiabatic transmission condition for an electron starting at the position $z_\text{S}$ with a magnetic field $B_\text{S} = B(z_\text{S})$, an electrostatic potential $U_\text{S} = U(z_\text{S})$, a kinetic energy $E = E(z_\text{S})$ with a corresponding gamma factor $\gamma$, and a pitch angle $\theta = \theta(z_\text{S})$.
The transmission condition then reads for all longitudinal positions $z$:
\begin{align}
        \nonumber
    0 & \leq \Elong(z)  \\
        \nonumber
      & = E + q U_\text{S} - \Etrans(z) - q U(z) \\
      & = E + q U_\text{S} - E \; \sin^2 \theta \cdot \frac{B(z)}{B_\text{S}} \; \frac{\gamma\!+\!1}{\gamma(z)\!+\!1} - qU(z) \; ,
    \label{eq:response:trans_condition_full}
\end{align}
where $\gamma(z)$ corresponds to the gamma factor at an arbitrary position $z$ along the beam line where the electron has a kinetic energy $E(z) = \Elong(z) + \Etrans(z)$ at a magnetic field $B(z)$ and an electrostatic potential $U(z)$.

Usually in a MAC-E filter the highest retarding potential $U$ and at the same time the smallest magnetic field $B_\text{A}$ is reached in the analyzing plane (located at $z_\text{ap} = 0$ in our definition). Secondly we can assume the electrical potential $U_\text{S}$ at the start to be zero and the relativistic factor in the analyzing plane at the largest retardation (minimum kinetic energy) to equal one, $\gamma(z_\text{ap}) = 1$.
Therefore the transmission condition in \cref{eq:response:trans_condition_full} simplifies to
\begin{equation}
    0 \leq E - E \; \sin^2 \theta \cdot \frac{B_\text{A}}{B_\text{S}} \; \frac{\gamma\!+\!1}{2} - qU \; .
    \label{eq:response:trans_condition}
\end{equation}
For a given electric potential and magnetic field configuration of the MAC-E filter, the transmission condition $\Tcal$ is thus just governed by the starting energy $E$, the starting angle $\theta$ and the retarding voltage $U$.
\begin{equation}
    \Tcal(E,\theta,U) =\left\{ \begin{array}{ll}
        1 &\text{if}\quad \displaystyle E \, \left(1 - \sin^2 \theta \cdot \frac{B_\text{A}}{B_\text{S}} \cdot \frac{\gamma\!+\!1}{2} \right) \\
        &\qquad\qquad - qU > 0\\
        0 &\text{else}
    \end{array}
    \right. \, .
    \label{eq:response:trans_condition_function}
\end{equation}
For an isotropically emitting electron source with angular distribution $\omega(\theta) \, \td \theta = \sin \theta  \, \td \theta$, we can integrate $\Tcal(E,\theta,U)$ over the angle $\theta$ and define a response or transmission function. From here on we associate the remaining energy in the analyzing plane of the MAC-E filter -- the \emph{surplus energy} -- with the expression $\eps = E - qU$.

In the KATRIN setup the maximum magnetic field $B_\text{max}$ is larger than $B_\text{S}$, so that \tbd electrons emitted at large pitch angles in the source are reflected magnetically before reaching the detector. The magnetic reflection occurs at the pinch magnet (with $B = B_\text{max}$ and zero potential), and in the source the electric potential is zero. The maximum pitch angle of the transmitted electrons is therefore independent of the electron energy and given by:
\begin{equation}
    \theta_\text{max} = \text{arcsin}\left(\sqrt{\frac{B_\text{S}}{B_\text{max}}}\right) \; ,
    \label{eq:response:theta_max}
\end{equation}
For the standard operating parameters of KATRIN (see \cref{tab:design}), $\theta_\text{max}$ evaluates to about $\SI{50.8}{\degree}$. This reflection is desired by design, since \tbd electrons emitted with larger pitch angles have to traverse a longer effective column of source gas and are therefore more likely to scatter and undergo energy loss, as detailed in the following sections.

\begin{figure}[ht]
    \centering
    \includegraphics[width=\figurewidthonecol]{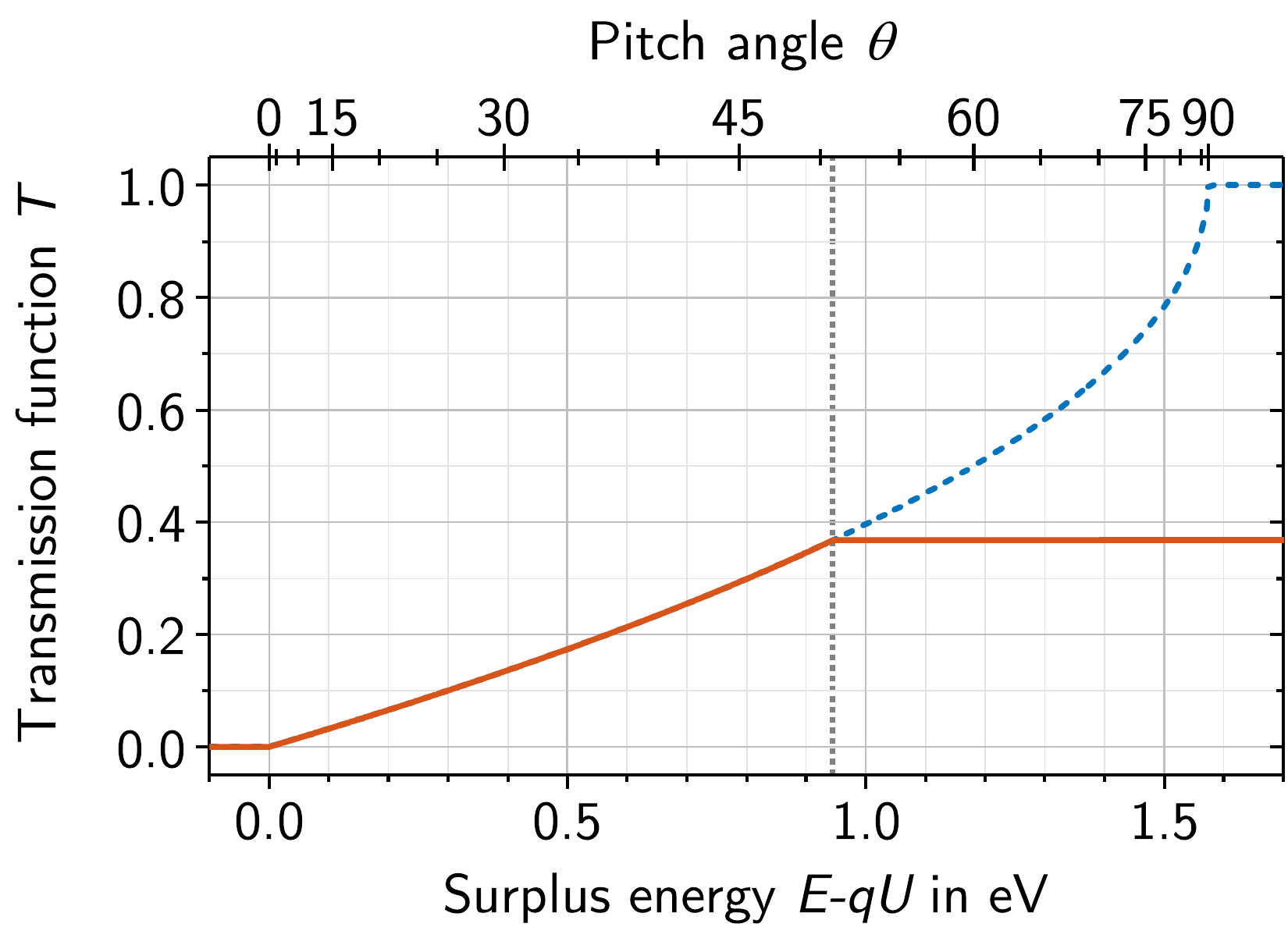}
    \caption[Transmission Function]{%
    Transmission function $T$ at a retarding potential of $U = \SI{18545}{V}$ with nominal magnetic field configuration ($\frac{B_\text{max}}{B_\text{A}} = \num{20000}$).
    The transmission condition in \cref{eq:response:trans_condition_function} relates the surplus energy to the pitch angle $\theta$, as shown at the top of the figure.
    The solid red line shows the cut-off caused by a magnetic reflection of all electrons with high pitch angle in the strongest magnetic field at reference conditions $\frac{B_\text{max}}{B_\text{S}} = \frac{6.0}{3.6}$.
    The dashed blue line shows the transmission function without magnetic reflection.
    }
    \label{fig:transmission}
\end{figure}

With this additional magnetic reflection after the analyzing plane, the transmission function is given by:
\begin{align}
        \nonumber
    T(E,U) &= \int_{\theta=0}^{\theta_\text{max}} \; \Tcal(E,\theta,U) \cdot \sin \theta \, \td \theta \\
      & = \left\{\begin{array}{ll}
        0 &\+ \eps<0 \\
        1 - \sqrt{1-\frac{\eps}{E} \frac{B_\text{S}}{B_\text{A}} \frac{2}{\gamma\!+\!1}} \, &\+ 0\leq \eps\leq \Delta E \\
        1 - \sqrt{1-\frac{B_\text{S}}{B_\text{max}}} &\+ \eps>\Delta E
    \end{array}
    \right. \, ,
    \label{eq:response:tf_simple}
\end{align}
with the filter width $\Delta E$ from \cref{eq:FilterWidth}. In \cref{fig:transmission}, the transmission function is shown for the nominal KATRIN operating parameters and for the case $B_\text{S} = B_\text{max}$.
The magnetic reflection imposes an upper limit on the pitch angle, which reduces the effective width of the transmission function. As indicated in \cref{fig:transmission}, this improves the filter width of the spectrometer to \SI{0.93}{eV}, compared with \SI{1.55}{eV} for $\theta_\mathrm{max} = \SI{90}{\degree}$ without magnetic reflection.

\subsection{Response function and energy loss}
\label{sec:eloss}

In the next step we consider the energy loss when the electron traverses the gaseous source. The dominant energy loss process is the scattering of electrons on gas molecules within the source. Because the pressure decreases rapidly outside the source, scattering processes in the transport section or thereafter are of no concern.

Two ingredients are required to appropriately treat electron scattering in the source.
First, the energy loss function $\tilde{f}(\epsilon, \delta \vartheta)$ describes the probability for a certain energy loss $\epsilon$ and scattering angle $\delta \vartheta$ of the \tbd electrons to occur in a scattering process.
Because the scattering angles $\delta \vartheta$ are small%
    \footnote{As investigated in~\cite{PhDGroh2015}, the direct angular change of \tbd electrons due to elastic and inelastic scattering has only negligible effect on the response function shape.}%
, we will neglect them in the following formulae and describe the scattering energy losses by the function $f(\epsilon)$.
Here we do not consider a dependence of $f$ or $P_s$ on the incident kinetic energy $E$ of the electrons, since for the KATRIN experiment the energy range of interest amounts to a very narrow interval of a few times \SI{10}{eV} below the tritium endpoint only, where these functions can be considered as independent of $E$.
The other important ingredients are the scattering probability functions $P_s(\theta)$ for an electron with pitch angle $\theta$ to scatter $s$ times before leaving the source. These scattering probabilities depend on $\theta$, since electrons with a larger pitch angle must traverse a longer path, meaning a larger effective column density, and are thus likely to scatter more often.

With these considerations, the response function no longer comprises only the transmission function, but is modified as follows:
\begin{align}
        \nonumber
    R(E, U) &= \int_{\epsilon = 0}^{E-qU} \; \int_{\theta=0}^{\theta_\text{max}} \; \Tcal(E-\epsilon,\theta,U) \cdot \sin \theta \\
        \nonumber
    &\qquad \cdot \biggl[ \; P_0(\theta) \, \delta(\epsilon) \, + \, P_1(\theta) \; f(\epsilon) \; \biggr. \\
    &\qquad\quad \biggl. \; + \, P_2(\theta) \; (f \otimes f)(\epsilon) \, + \, \ldots \; \biggr] \, \td \theta \, \td \epsilon \\
        \nonumber
    &= \int_{\epsilon = 0}^{E-qU} \; \int_{\theta=0}^{\theta_\text{max}} \; \Tcal(E-\epsilon,\theta,U) \cdot \sin \theta \\
    &\qquad \cdot \, \sum_{s} \; P_s(\theta) \; f_s(\epsilon) \, \td \theta \, \td \epsilon \; .
    \label{eq:simulation:theo:response}
\end{align}
Electrons leaving the source without scattering $(s = 0)$ do not lose any energy, hence $f_0(\epsilon) = \delta(\epsilon)$. For $s$-fold scattering, $f_s(\epsilon)$ is obtained by convolving the energy loss function $f(\epsilon)$ $s$ times with itself.

The scattering cross section can be divided into an elastic and an inelastic component.
The inelastic cross section and the energy loss function for electrons with kinetic energies of $\approx \SI{18.6}{keV}$ scattering from tritium molecules have both been measured in~\cite{Aseev2000,bib:abdurashitov}. In this work, the inelastic scattering cross section was determined to be $\sigma_\text{inel} = \SI{3.40(7)E-18}{cm^2}$ and an empirical model was fit to the energy loss spectrum.

The latter is parameterized by a low-energy Gaussian and a high-energy Lorentzian part:
\begin{equation}
    f(\epsilon) = \left\{\begin{array}{ll}
        A_1 \cdot \, \exp\left(-2 \, \left(\dfrac{\epsilon - \epsilon_1}{\omega_1}\right)^2 \right) &\+ \epsilon < \epsilon_c \\
        A_2 \cdot \, \dfrac{\omega_2^2}{\omega_2^2 + 4 (\epsilon - \epsilon_2)^2} &\+ \epsilon \geq \epsilon_c \\
    \end{array}
    \right. \, ,
    \label{eq:eloss_aseev}
\end{equation}
with $A_1 = (0.204 \pm 0.001)\,\text{eV}^{-1}$, $A_2 = (0.0556 \pm 0.0003)\,\text{eV}^{-1}$, $\omega_1 = (1.85 \pm 0.02)\,\text{eV}$, $\omega_2 = (12.5 \pm 0.1)\,\text{eV}$, $\epsilon_2 = (14.30 \pm 0.02)\,\text{eV}$ and a fixed $\epsilon_1 = 12.6\,\text{eV}$. To obtain a continuous transition between the two parts of $f(\epsilon)$, a value $\epsilon_c = 14.09\,\text{eV}$ was chosen. The Gaussian part summarizes the energy loss due to (discrete) excitation processes, while the Lorentzian part describes the energy loss due to ionization of tritium molecules.

This parameterization of the energy loss function is used for the response model presented in this paper. However, the parameters are not precise enough for KATRIN to meet its physics goals. Dedicated electron gun measurements with the full experimental KATRIN setup have been planned for the determination of the inelastic scattering cross section and the energy loss function with higher precision; the analysis of these data will involve a sophisticated deconvolution technique~\cite{Hannen2017}.

\begin{figure}[ht!]
	\centering
	\begin{subfigure}{\linewidth}
		\centering
		\includegraphics[width=\linewidth]{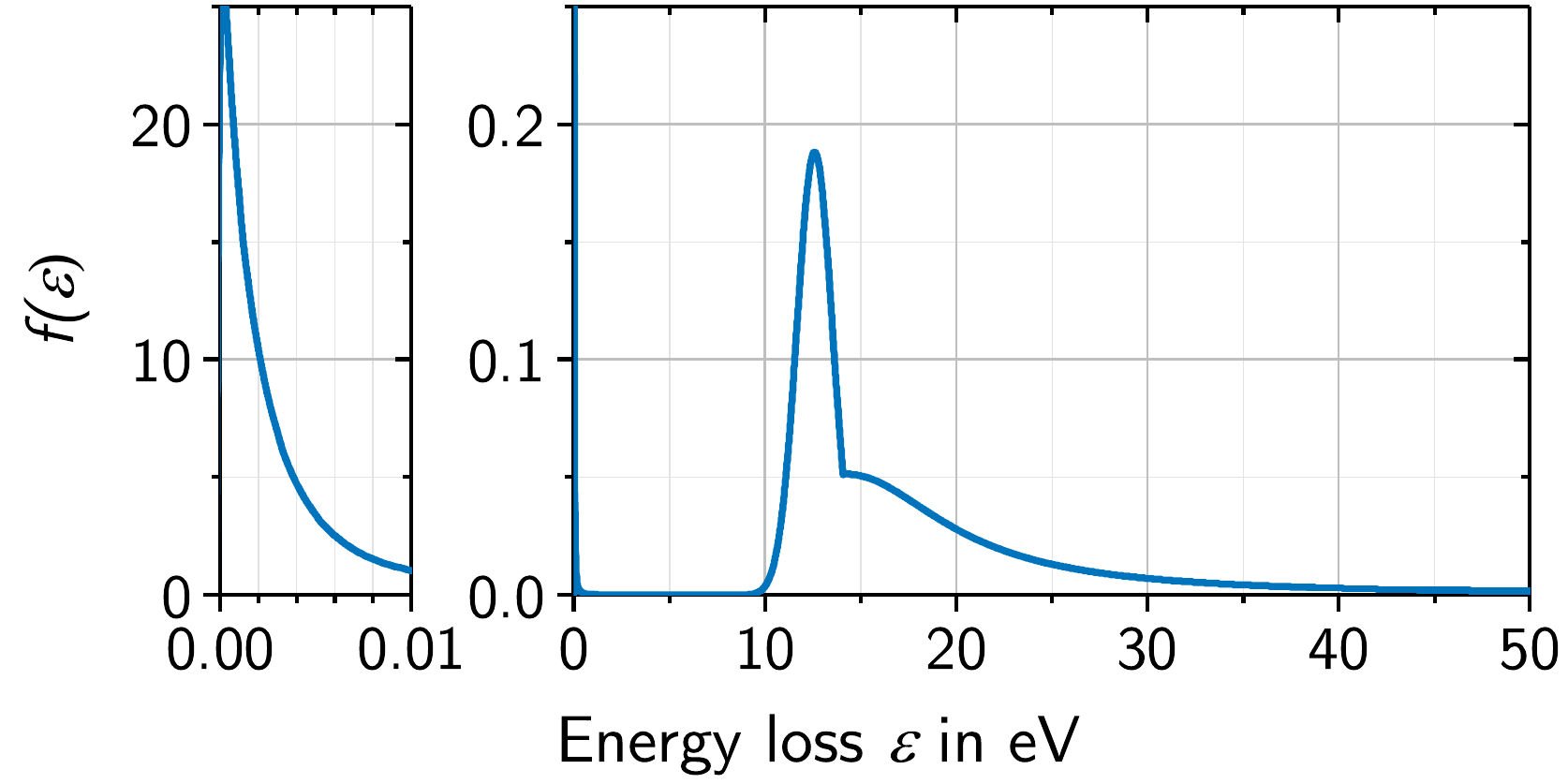}
		\subcaption{Energy loss function $f_1 = f(\epsilon)$ --- the energy loss probability of electrons scattered once. Shown is the normalized probability distribution, $\int_0^\infty f_1(\epsilon) \td \epsilon = 1$.}
	\end{subfigure}
	\par\bigskip
	\begin{subfigure}{\linewidth}
		\centering
		\includegraphics[width=\linewidth]{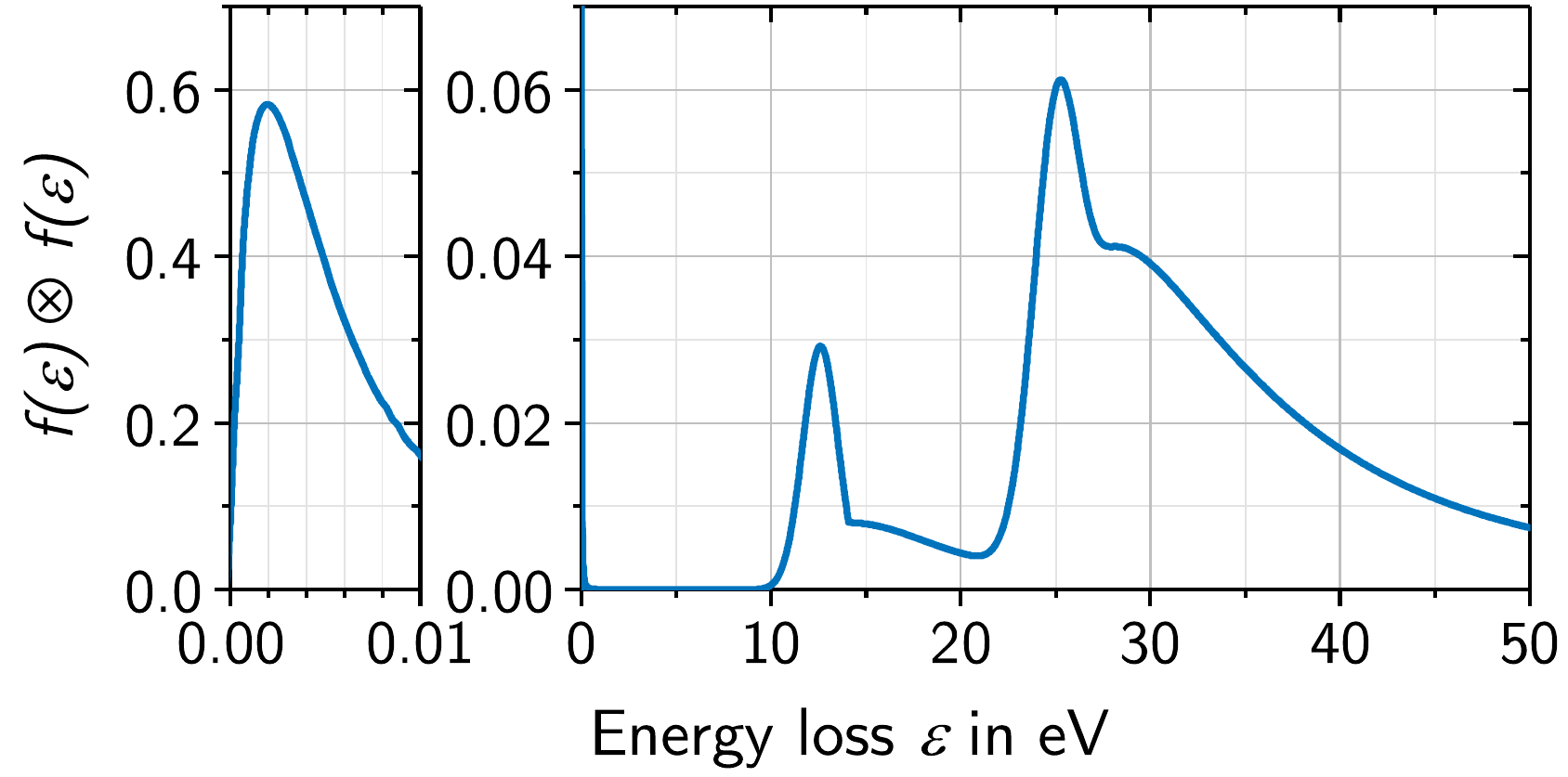}
		\subcaption{Convolved energy loss function $f_2 = f(\epsilon) \otimes f(\epsilon)$ --- the energy loss probability of electrons scattered twice.}
	\end{subfigure}
	\caption[Energy Loss Function]{%
		Theoretical energy loss function for elastic and inelastic scattering processes, shown as a probability density function. The leftmost enlarged region ($\epsilon \lesssim \SI{0.01}{eV}$) is dominated by elastic scattering, and the region at higher energy is due to inelastic excitation and ionization, as parameterized by Aseev et al.~\cite{Aseev2000}.}
	\label{fig:eloss}
\end{figure}

At $\sigma_\mathrm{el} = 0.29 \cdot 10^{-18}\,\mathrm{cm}^{2}$, the total cross section of elastic scattering of \SI{18.6}{keV} electrons with molecular hydrogen isotopologues is smaller than that for inelastic scattering by an order of magnitude~\cite{bib:geiger,bib:liu2}.
In addition, the elastically scattered electrons are strongly forward peaked with a median scattering angle of $\overline\theta_\text{scat} = \SI{2.1}{\degree}$ near the tritium endpoint energy.
The energy loss due to elastic scattering is given by the relation
\begin{equation}
    \Delta E_\text{scat} = 2 \; \frac{\me}{M_\chem{T_2}} \; E \cdot \left( 1-\cos \theta_\text{scat} \right) \; .
\end{equation}
With an angular distribution for elastic scattering of molecular hydrogen by electron impact measured in ~\cite{bib:nis}, the corresponding median energy loss amounts to $\overline{\Delta E} = \SI{4.0}{meV}$. The energy loss function, containing the elastic and inelastic components weighted by their individual cross section, is shown in \cref{fig:eloss}.

The elastic energy loss component can be accurately calculated. Due to its narrow width and steep slope, $\sim \si{meV}$ binning is required for incorporating it accurately in the response function, thereby increasing computational cost considerably. We will neglect the elastic scattering component in neutrino mass measurements as the associated systematic error on an $\mnusq$ is minute ($\sim \SI{5E-5}{eV^2}$, see \cref{tab:systematics}).

\subsection{Radial inhomogeneity of the electromagnetic field}
\label{sec:transmission_inhomogeneity}

To calculate the transmission and response functions of the KATRIN setup as explained in \cref{sec:transmission} and \cref{sec:eloss}, it is in principle sufficient to only consider the axial position of an electron to identify the initial conditions such as electromagnetic fields or scattering probabilities.
In the case of the main spectrometer, radial dependencies must be incorporated in the description of the magnetic field and the electrostatic potential in the analyzing plane. Additional radial dependencies in the source are discussed in \cref{sec:source}; these are then incorporated into the model together with the spectrometer effects.

In order to achieve a MAC-E filter width in the eV-regime, a reduction of the magnetic field strength in the analyzing plane on the order of $\frac{B_\text{A}}{B_\text{max}} \approx \frac{\Delta E}{E} \approx \num{e-4}$ is required (see \cref{eq:FilterWidth}).
Consequently the diameter of the flux-tube area $A$ is drastically increased due to the conservation of magnetic flux $\Phi = \mathrm{const} \approx B \cdot A$.
When nominal field settings are applied (see \cref{tab:design}), the projection of the detector surface with radius $r_\mathrm{det} = \SI{4.5}{cm}$ has a radius of about \SI{4}{m} in the analyzing plane.
A larger (smaller) magnetic field in the analyzing plane $B_\mathrm{A}$ shifts the transmission edge to a larger (lower) energy, see \cref{eq:response:trans_condition_function}.
This effect is even more pronounced for larger electron pitch angles.
Consequently, the transmission function (see \cref{eq:response:tf_simple}) is also widened or narrowed.
Utilizing a set of magnetic field compensation coils, operated with an optimal current distribution, around the spectrometer vessel, the spread of the radial inhomogeneity of the magnetic field is minimized to a few \si{\micro T} when an optimized current distribution is applied~\cite{Glueck2013,Erhard2018}.
The resulting variation in the filter width in the analyzing plane due to the magnetic field inhomogeneity is thus reduced to about \SI{10}{meV}~\cite{PhDErhard2016}.

In the case of the electrostatic potential, unavoidable radial variation arises from the design of the spectrometer. To fulfill the transmission condition in \cref{eq:response:trans_condition}, the electrode segments at the entrance and exit are operated on a more positive potential than in the central region close to the analyzing plane%
    \footnote{It is required that $E_\parallel$ reaches its global minimum in the analyzing plane, which is achieved by optimizing the electromagnetic conditions in the spectrometer. See \cite{Glueck2013} for details.}%
.
Depending on the final potential setting, the radial potential variation in the analyzing plane is expected to be of order \SI{1}{V}~\cite{PhDGroh2015}. In comparison, azimuthal variations are negligible. It is possible to considerably reduce the radial potential inhomogeneity by operating the MAC-E filter at larger $B_\mathrm{A}$. However, this would require better knowledge of the magnetic field in the analyzing plane~\cite{PhDErhard2016} and also increase the filter width.

\begin{figure}[ht]
    \centering
    \includegraphics[width=\figurewidthonecol]{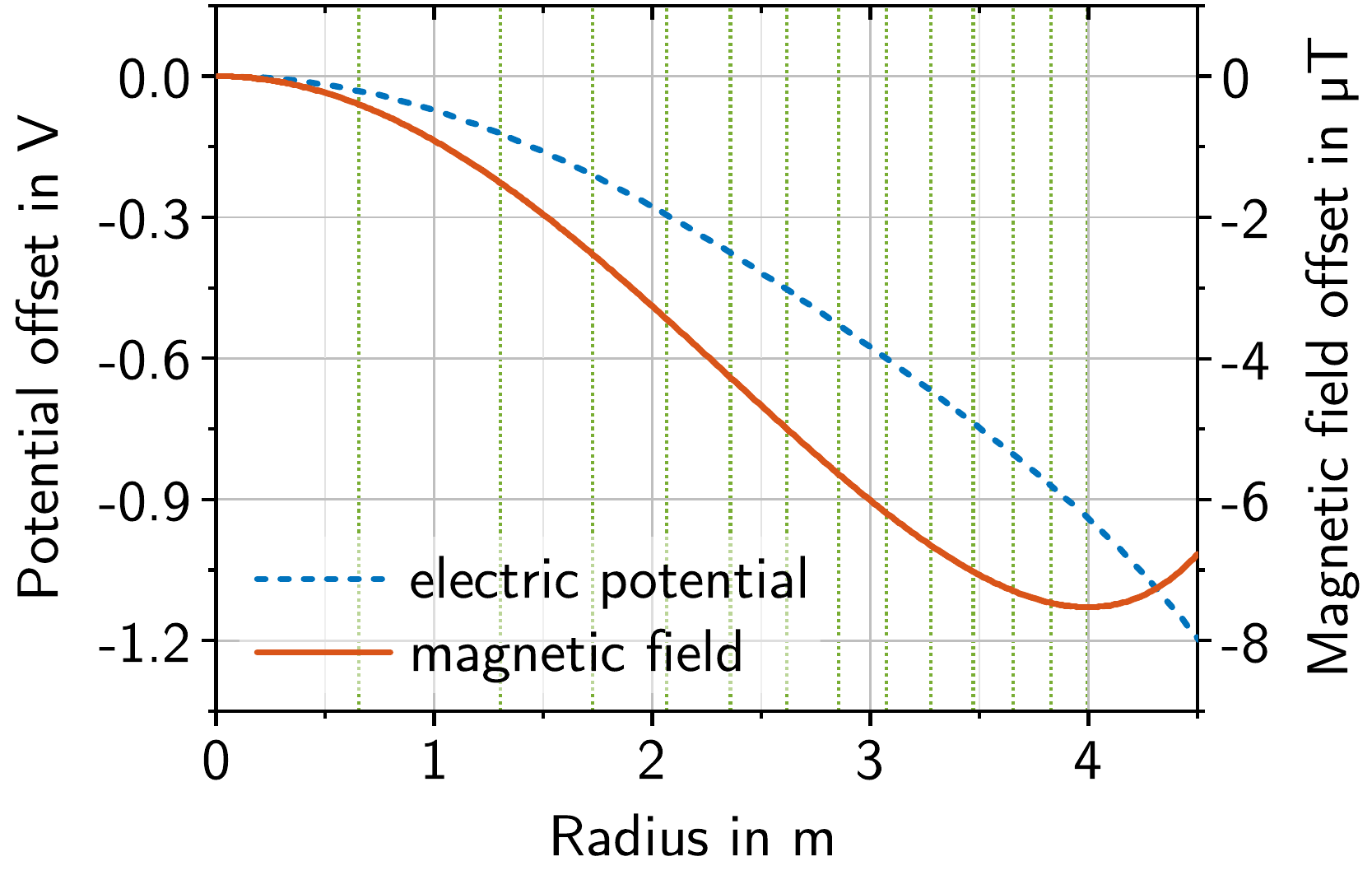}
    \caption{%
    The calculated radial inhomogeneity of the electrostatic potential and the magnetic field in the analyzing plane of the main spectrometer, for the standard setting of $U = \SI{-18600}{V}$ and $B_\text{A} = \SI{0.3}{mT}$.
    The plot shows the offset in the potential and the magnetic field values in the spectrometer center. The vertical dashed lines mark the corresponding outer radii of annuli mapped to the \num{13} detector rings.}
    \label{fig:radial_TF}
\end{figure}

Even with these optimizations of the setup, the small radial variations in the electromagnetic fields at the analyzing plane, as shown in \cref{fig:radial_TF}, cannot be neglected.
The segmentation of the KATRIN main detector into annuli of pixels allows us to incorporate such radial variations in the response function model for each individual detector pixel. Because the tritium source also features radial variations of certain parameters, this segmentation is combined with a full segmentation of the source volume as described in \cref{sec:source}.
Dependencies of the electromagnetic field are typically averaged over the surface area of a pixel. The specific detector geometry with thinner annuli towards outer radii (each with equal surface area) helps minimize the potential variation within individual annuli, despite the increasing steepness of the potential.

\subsection{Source volume segmentation and effects}
\label{sec:source}

\begin{figure*}[htb]
	\centering
	\includegraphics[width=0.92\figurewidthtwocols]{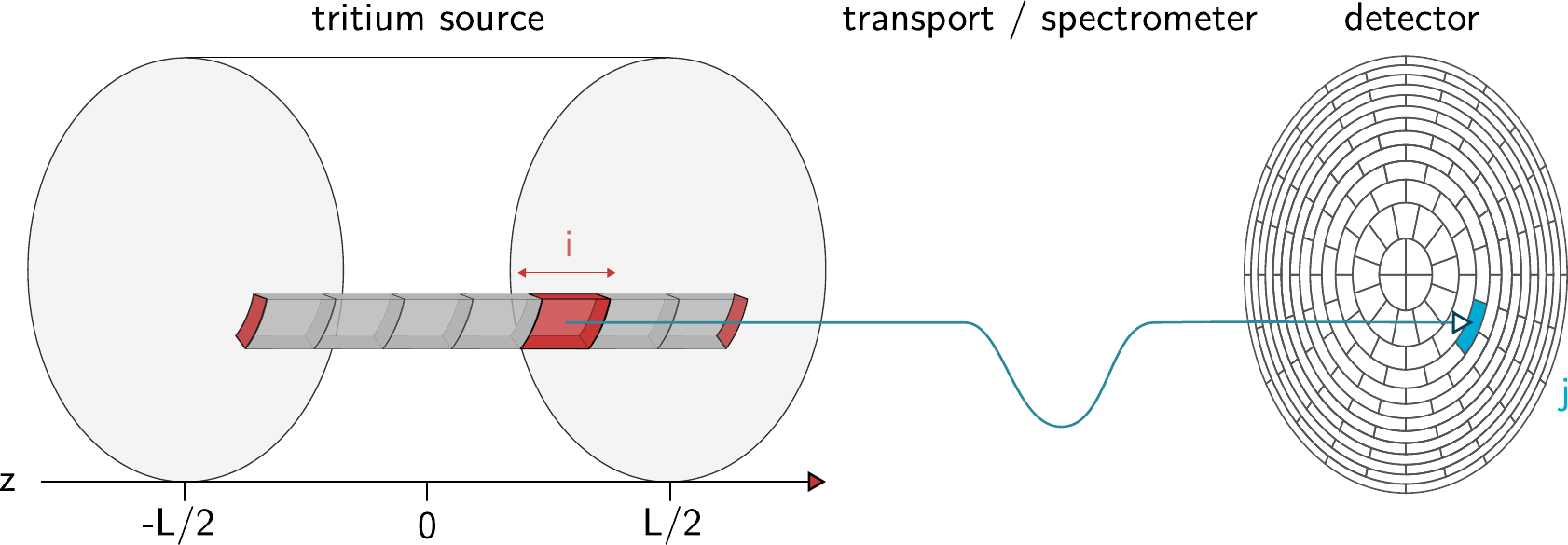}
	\caption[Source Segmentation]{%
		In the numerical model, the source is partitioned in such a way that each radial/azimuthal segment (index $j$) in the source, consisting of stacked longitudinal slices (index $i$), corresponds to the part of the magnetic flux tube seen by the matching detector pixel (index $j$). (Diagram not drawn to scale.)
	}
	\label{fig:seg}
\end{figure*}

In addition to radial dependencies of the analyzing plane parameters that govern the energy analysis of the \tbd electrons (\cref{sec:transmission_inhomogeneity}), the tritium source also features radial and axial dependencies of its parameters.
In the following, we will briefly outline the most relevant source parameters that are required to accurately model the differential \tb spectrum and the response function. These parameters include the beam tube temperature $T_\text{bt}$, the magnetic field strength $B_\text{S}$, plasma potentials $U_\text{P}$, the particle density $\rho$ and the bulk velocity $u$ of the gas, all of which may vary slightly in longitudinal, radial and azimuthal directions.
The complex gas dynamic simulations, which are needed to calculate these local source parameters, are described in comprehensive detail in~\cite{PhDKuckert2016,arXivKuckert2018}.

In order to model accurately these effects for each individual detector pixel, the simulation source model is partitioned to match the detector geometry. It is partitioned longitudinally into $N_L$ slices and segmented radially into $N_R$ annuli (rings) of $N_S$ segments each, resulting in a total of $N_L \cdot N_R \cdot N_S$ segments (see \cref{fig:seg}). The geometry of these segments is chosen in such a way, that a longitudinal stack of segments is magnetically projected%
    \footnote{The \tbd electrons are guided from source to detector by magnetic field lines, so each detector pixel maps a certain stack of source segments.}
onto a corresponding detector pixel. Note that all detector pixels have identical surface area, which leads to broader annuli at the center and thinner annuli towards larger radii. In the following, we index the longitudinal slices by the subscript $i$ and radial/azimuthal segments with their corresponding detector pixel by the subscript $j$.

At a retarding potential $U$, the detection rate for a specific detector pixel $j$ can then be stated as
\begin{equation}
    \dot N_j(U) = \frac{1}{2} \, \sum_{i=0}^{N_L-1} \; N_{\text{T},i} \, \int_{qU}^{E_0} \; \frac{\td \Gamma}{\td E}(E_0, m_\upnu^2) \; R_{i,j}(E, U) \, \td E \; ,
    \label{eq:sliced_rate}
\end{equation}
where $N_{\text{T},i}$ is the number of tritium nuclei (assuming that the gas density has no radial or azimuthal dependence).
The response function $R_{i,j}(E, U)$ depends on the index $i$ (i.e.~the axial position) and the index $j$ (i.e.~the radial/azimuthal position) of the source segment. With the indices $i,j$ we can describe the dependence on local source parameters such as the magnetic field. The most significant effect on the response is caused by the scattering probabilities, as detailed in \cref{sec:eloss}.
The index $j$ further describes non-uniformities of the retarding potential $U$ and the magnetic field $B_\text{A}$ in the spectrometer (see \cref{fig:radial_TF}).

\subsection{Scattering probabilities}
\label{sec:scattering}

\begin{figure}[ht]
    \centering
    \includegraphics[width=\figurewidthonecol]{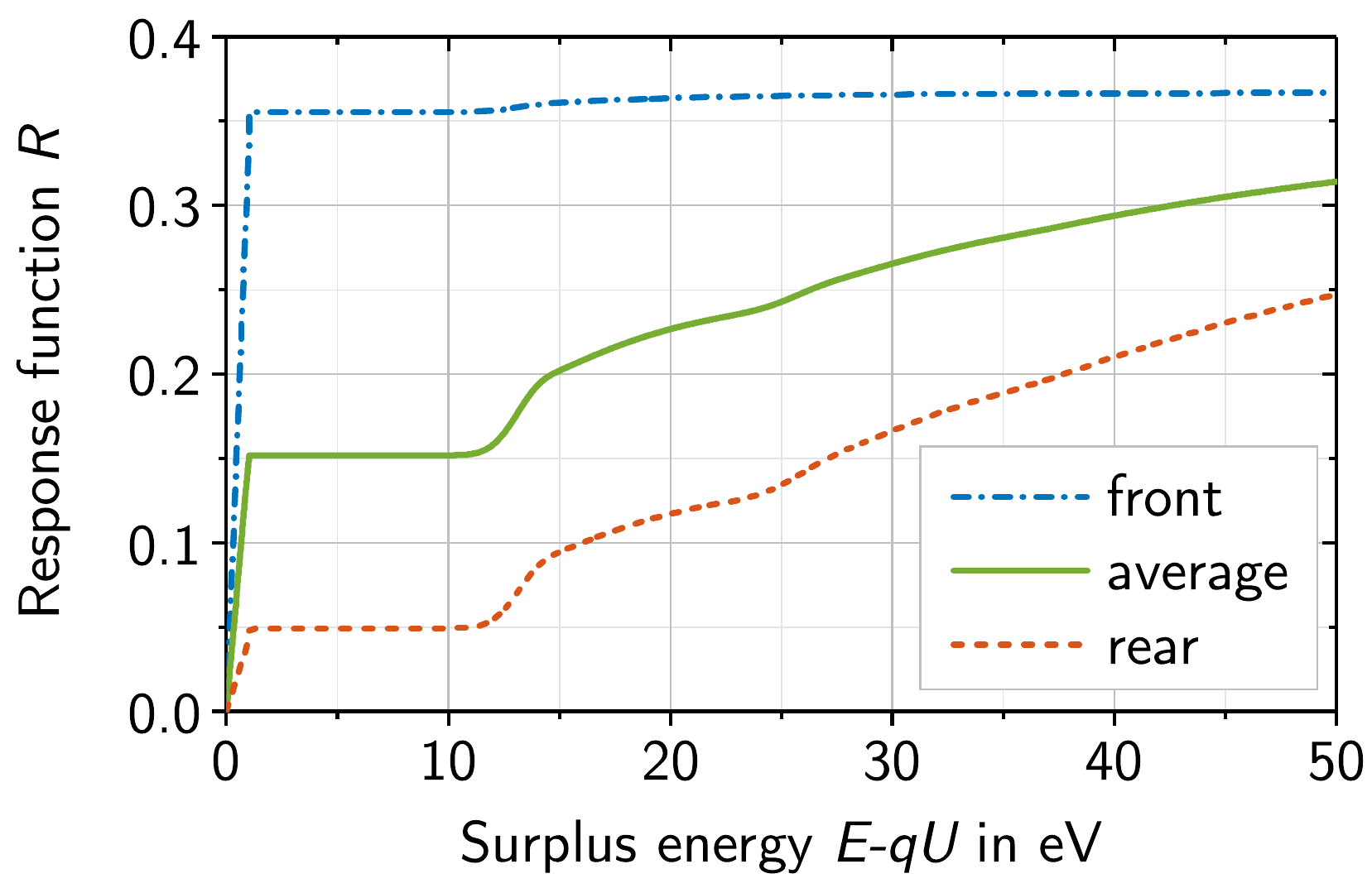}
    \caption[Response Function]{%
    The response function $R(E, qU)$ at a retarding energy of $qU = \SI{18545}{eV}$. The dash-dotted and dashed curves show the response function close to the front (spectrometer-facing, $z=\SI{+4}{m}$) vs.\ rear ($z=\SI{-4}{m}$) of the WGTS, which has a length of \SI{10}{m} in total. An averaged version, weighted by the gas density in each source segment, is shown as the solid curve.
    }
    \label{fig:response}
\end{figure}

As discussed in \cref{sec:eloss}, inelastic scattering results in an energy loss that directly affects the energy analysis of the signal electrons, and needs to be incorporated accurately into the analytical description. Changes to the angular distribution of the emitted electrons due to scattering processes, which also modify the response function, are discussed in \cref{sec:response_noniso}.

The scattering probability for \tbd electrons is considerably different depending on their starting position in the \SI{10}{m} long source beam tube, as visualized in \cref{fig:response}. The longitudinal segmentation of the source volume in our model allows us to incorporate this behavior.
The probability $P_s$ for an electron to leave the source after scattering exactly $s$ times depends on the total cross section $\sigma$ and the effective column density $\Ncal_\text{eff}$ that the electron traverses. This effective column density depends not only on the electron's starting position $z$ inside the source and the axial density distribution $\rho(z)$, but also on the starting pitch angle $\theta$ in the source (\cref{eq:response:theta_max}):
\begin{equation}
    \Ncal_\text{eff}(z,\theta) = \frac{1}{\cos(\theta)} \cdot \int_{z}^{L/2} \; \rho(z') \, \td z' \; .
    \label{eq:scat_prob1}
\end{equation}
$L$ denotes the length of the source beam tube with $-L/2 \le z \le L/2$. The nominal column density is then given by $\Ncal = \Ncal_\text{eff}(z=-L/2,\ \theta=0)$.

Because of the low probability to scatter off a single tritium molecule, the number of scatterings during propagation can be calculated according to a Poisson distribution:
\begin{equation}
    P_s(z, \theta) = \frac{(\, \Ncal_\text{eff}(z,\theta) \cdot \sigma \,)^s}{s!} \, \cdot \, \exp( -\Ncal_\text{eff}(z,\theta)\cdot\sigma ) \; .
    \label{eq:scat_prob2}
\end{equation}
The mean scattering probabilities for a specific position $z$ can be calculated using the isotropic angular distribution $\omega(\theta) = \sin{\theta}$ and the maximum pitch angle $\theta_\text{max}$:
\begin{equation}
    P_s(z) = \frac{1}{1-\cos(\theta_\mathrm{max})} \; \int_{\theta=0}^{\theta_\mathrm{max}} \; \sin(\theta) \; P_s(z,\theta) \, \td \theta \; .
    \label{eq:scat_prob3}
\end{equation}
This integration assumes that the angular distribution is not significantly affected by the small angular change in the discussed scattering processes. A higher total column density $\Ncal$, as well as a larger $\theta_\text{max}$, would provide a larger number of \tbd electrons at the exit of the source and at the detector. However, they also raise the proportion of scattered over unscattered electrons, thereby increasing the systematic uncertainties due to energy loss, and at some point, limiting the \tbd electron detection rate close to the endpoint. The optimal design values of $\Ncal = \SI{5e17}{cm^{-2}}$ and $ \theta_\text{max} = \SI{50.8}{\degree}$~\cite{KATRIN2005} balance these effects.

\subsection{Response function for non-scattered electrons}
\label{sec:response_noniso}

The transmission function in \cref{eq:response:tf_simple} describes the transmission probability of isotropically emitted electrons. Even if we consider only non-scattered electrons, the \tbd electrons do not follow an isotropic angular distribution before entering the spectrometer due to the pitch angle dependence of the $s$-fold scattering probabilities $P_s(z,\theta)$ in the source (see \cref{sec:scattering}).

\begin{figure}[ht]
    \centering
    \includegraphics[width=\figurewidthonecol]{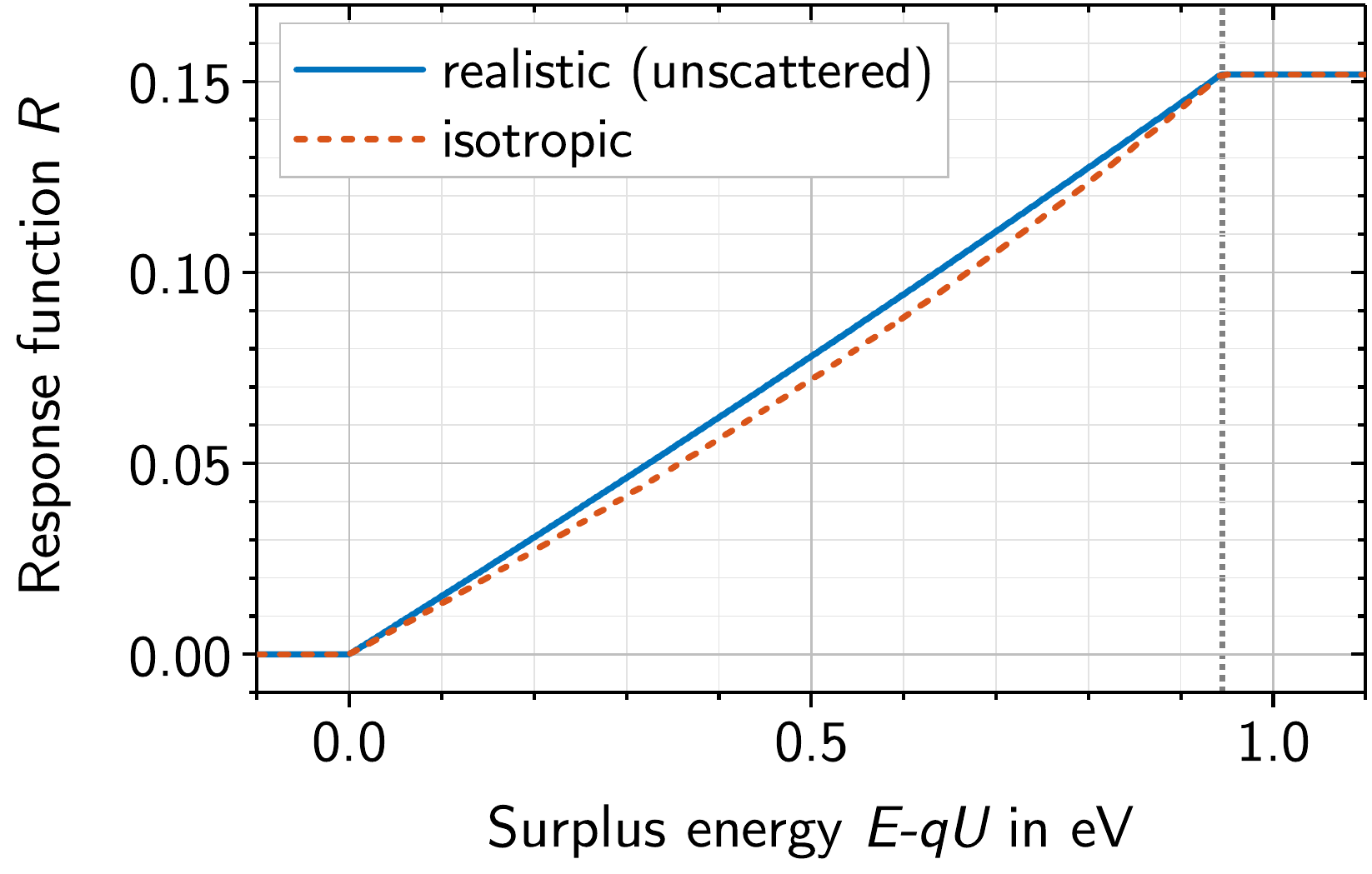}
    \caption[Realistic transmission edge]{%
    The transmission edge of the response function. The dashed curve is calculated with an isotropic angular distribution, and the solid curve with a realistic angular distribution for unscattered electrons.}
    \label{fig:transmission_non_iso}
\end{figure}

The zero-scattering transmission function therefore needs to be modified to the following form:
\begin{align}
        \nonumber
    T^\star_s(E, U) &= \biggl. R(E,U) \biggr|_{\; \eps \, < \, \SI{10}{eV}} \\
      &= \int_{\theta=0}^{\theta_\text{max}} \; \Tcal(E,\theta,U) \cdot \sin \theta \; P_0(\theta) \, \td \theta \; .
    \label{eq:tf_noniso}
\end{align}
The zero-scattering probability $P_0(\theta)$ is computed by averaging $P_0(z,\theta)$ over $z$.
\Cref{fig:transmission_non_iso} illustrates the resulting difference in the response function. The surplus energy range $\eps < \SI{10}{eV}$ corresponds to the steep increase in the response function at low energies as shown in \cref{fig:response}, where energy loss from inelastic scattering does not contribute.

\subsection{Doppler effect}
\label{sec:doppler}
\index{Doppler effect}

The thermal translational motion and the bulk gas flow of the \tbd emitting tritium molecules in the WGTS lead to a Doppler broadening of the electron energy spectrum, which further modifies the response function model that was derived in \cref{sec:eloss} and thereafter.
These two effects can be expressed as a convolution of the differential spectrum $\frac{\td \Gamma}{\td E}$ with a broadening kernel $g$, denoted by the subscript $\text{D}$:
\begin{align}
    \left(\frac{\td \Gamma}{\td E}\right)_\text{D} &= \left( g \otimes \frac{\td \Gamma}{\td E} \right)(\Elab) \\
    &= \int_{-\infty}^{+\infty} \; g(\Ecms, \Elab) \; \frac{\td \Gamma}{\td E}(\Ecms) \, \td \Ecms
    \label{eq:doppler}
    \: ,
\end{align}
with $\Ecms$ being the electron kinetic energy in the \tbd emitter's rest frame (which is approximately the center-of-mass system), and $\Elab$ the electron energy in the laboratory frame.

The magnitude of the thermal tritium gas velocity follows a Maxwell-Boltz\-mann distribution. However, considering only the velocity component $\vM$ that is parallel to the electron emission direction, the thermal velocity distribution of the tritium isotopologue mass $M$ is described by a Gaussian
\begin{equation}
    \label{eq:thermal}
    g(\vM) = \frac{1}{\sqrt{2\pi}\sigma_v} \cdot \upe^{ -\frac{1}{2} \left(\frac{\vM}{\sigma_v}\right)^2 } \; ,
\end{equation}
which centers around $\vM = 0$ with a standard deviation $\sigma_v = \sqrt{\kB T_\text{bt} / M}$.
For the component of the bulk gas velocity $u$ that is parallel to the electron emission direction with pitch angle $\theta$, the mean $\vM$ is shifted by $\cos\theta \cdot u$. Integrating over all emission directions up to $\theta_\text{max}$, the expression expands to
\begin{align}
        \nonumber
    g(\vM) &= \frac{1}{(1-\cos\theta_\text{max})} \cdot  \\
      &\quad \cdot \! \int_{\cos\theta_\text{max}}^1 \! \frac{1}{\sqrt{2\pi}\sigma_v} \cdot \upe^{ -\frac{1}{2} \left(\frac{\vM - \cos\theta \cdot u}{\sigma_v}\right)^2 } \,  \td \cos\theta \; .
    \label{eq:doppler_comb}
\end{align}
Using the Gaussian error function this expression can be rewritten as
\begin{align}
        \nonumber
    g(\vM) &= \frac{1}{(1-\cos\theta_\text{max})\cdot 2u} \\
      &\quad \cdot \text{erf}\left(\frac{\vM - \cos\theta_\text{max} \cdot u}{\sqrt{2}\,\sigma_v}, \frac{\vM - u}{\sqrt{2}\,\sigma_v}\right) \; .
\end{align}
Finally, the tritium gas velocity distribution $g(\vM)$ can be translated into an electron energy distribution $g(\Ecms, \Elab)$. Using the Lorentz factors and the electron velocities defined in the CMS and lab frames, we can write
\begin{equation}
    g(\Ecms, \Elab) = \frac{g(\vM)}{\gcms \, \me \, \vecms}
\end{equation}
with
\begin{equation*}
    \vM \approx \frac{\velab - \vecms}{1 - \velab \cdot \vecms / c^2} \; .
\end{equation*}
The standard deviation of this convolution kernel evaluates to
\begin{align}
        \nonumber
    \sigma_E &= \sigma_v \; \gcms \; \me \; \vecms \\
    &= \sqrt{(\Ecms + 2 \me) \, \Ecms \cdot \kB T_\text{bt} / M} \; .
\end{align}

\begin{figure}[ht]
    \centering
    \includegraphics[width=\figurewidthonecol]{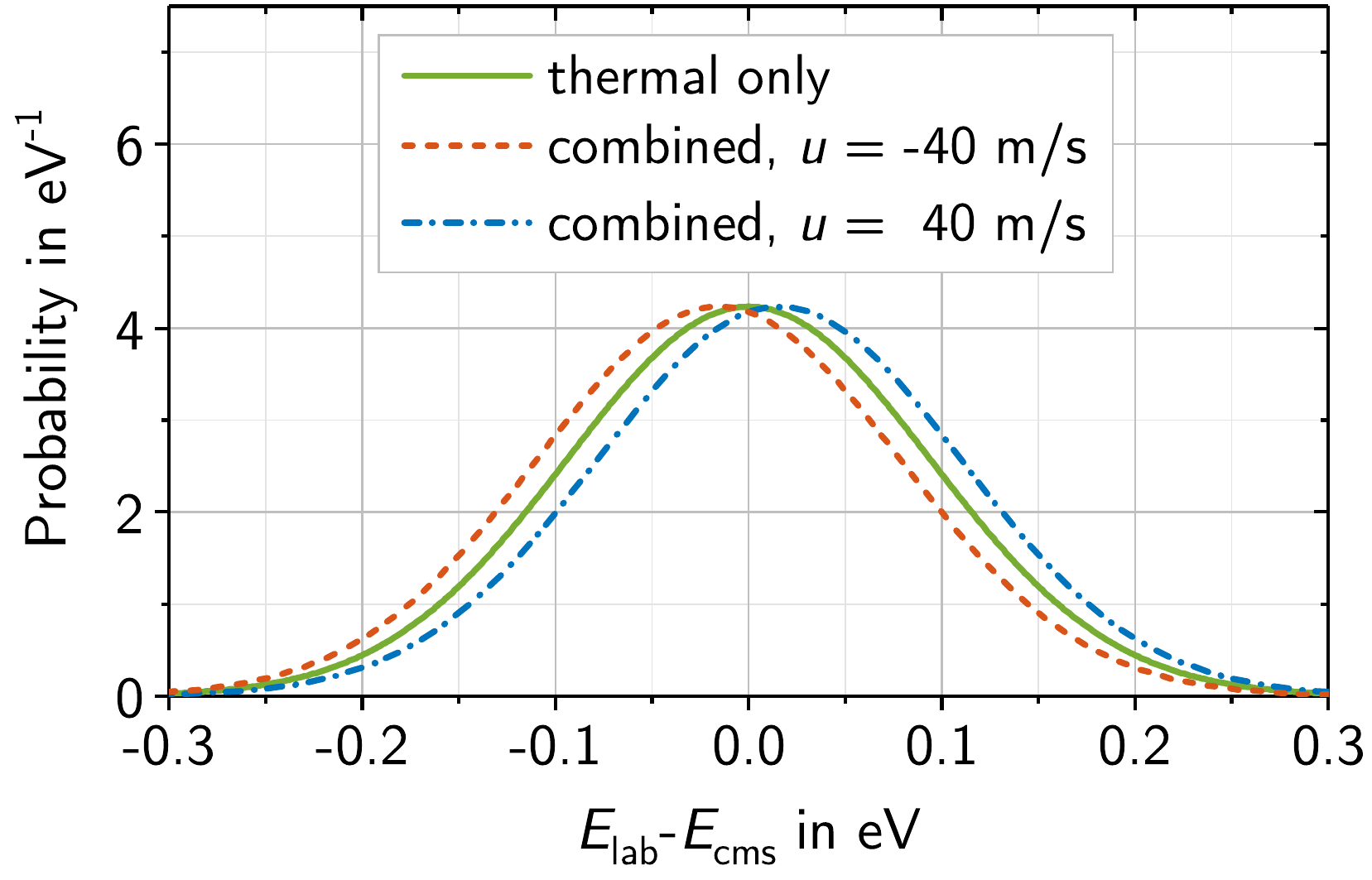}
    \caption[Doppler broadening kernel]{%
    Convolution kernels describing the Doppler broadening of the \tb spectrum due to the thermal motion and bulk velocity $u$ of the source gas. A temperature of $T_\text{bt} = \SI{30}{K}$ is assumed, leading to a Gaussian broadening with $\sigma_E \approx \SI{94}{meV}$ at $\Ecms = \SI{18575}{eV}$.}
    \label{fig:doppler}
\end{figure}

With $\sigma_v \approx \SI{203}{m/s}$ for $\chem{T_2}$ molecules at $T_\text{bt}=\SI{30}{K}$ and the weighted mean bulk velocity at nominal source conditions being $\bar{u} \approx \SI{13}{m/s}$, thermal Doppler broadening clearly is a dominating effect. The standard deviation of the broadening function $g(\Ecms, \Elab)$ at a fixed bulk velocity $u = 0$ for $T_\text{bt} = \SI{30}{K}$ and $E \approx E_0$ evaluates to $\sigma_E \approx \SI{94}{meV}$ (also see \cref{fig:doppler}). This value can be interpreted as a significant smearing of the energy scale. Its implication for the neutrino mass measurement is shown in \cref{tab:systematics}.

\subsection{Cyclotron radiation}
\label{sec:cyclotron}

As electrons move from the source to the spectrometer section in KATRIN, they lose energy through cyclotron radiation. In contrast to energy loss due to scattering with tritium gas (\cref{sec:scattering}), this energy loss process applies to the entire trajectory of an electron as it traverses the experimental beamline~\cite{Arenz2018b}.

For a particle with kinetic energy $E$ spending a time $\Delta t$ in a fixed magnetic field $B$, the cyclotron energy loss is (in SI units):
\begin{equation}
    \Delta E_\perp^\text{cycl} = -\frac{q^4}{3 \pi c^3 \varepsilon_0 \me^3} \cdot B^2 \cdot \Etrans \; \frac{\gamma\!+\!1}{2} \cdot \Delta t \; .
\end{equation}
In general, cyclotron radiation reduces the transverse momentum component of the particle%
    \footnote{In the non-relativistic case, the power loss due to cyclotron radiation amounts to $\dot{E}_\perp = -\SI{0.39}{/s.T^2} \cdot E_\perp \cdot B^2$.}%
. Consequently, the losses are maximal for large pitch angles and vanish completely at $\theta = \SI{0}{\degree}$.

For complex geometric and magnetic field configurations as in the KATRIN experiment, the overall cyclotron energy loss can be computed using a particle tracking simulation framework such as Kassiopeia~\cite{Furse2017}.
By this means, the cyclotron energy loss from the source to the analyzing point in the main spectrometer can be obtained as a function of the electron's starting position $z$ and pitch angle $\theta$. Particles starting in the rear of the source will lose more energy due to their longer path through the whole setup. The total cyclotron energy loss can be up to \SI{85}{meV} for electrons with the maximum pitch angle $\theta_\text{max} = \SI{50.8}{\degree}$.

Because the resulting decrease in the angle $\Delta \theta$ due to the loss of transverse momentum is of order \num{e-6} or less, it can be neglected. We thus consider the loss of cyclotron energy $\Delta E^\text{cycl}(\theta,z)$ to be a decrease in the total electron kinetic energy $E$. Essentially, this effect causes a shift of the electron transmission condition (see \cref{eq:response:trans_condition_function})
\begin{equation}
    \Tcal^\text{cycl}_i(E,\theta,U) = \Tcal(E - \Delta E^\text{cycl}(\theta,z),\theta,U)
    \label{eq:response:trans_condition_function_cyclotron}
\end{equation}
with the index $i$ denoting the longitudinal slice where the electron starts from the source position $z$ (see \cref{fig:seg}).

\begin{figure}[ht]
    \centering
    \includegraphics[width=\figurewidthonecol]{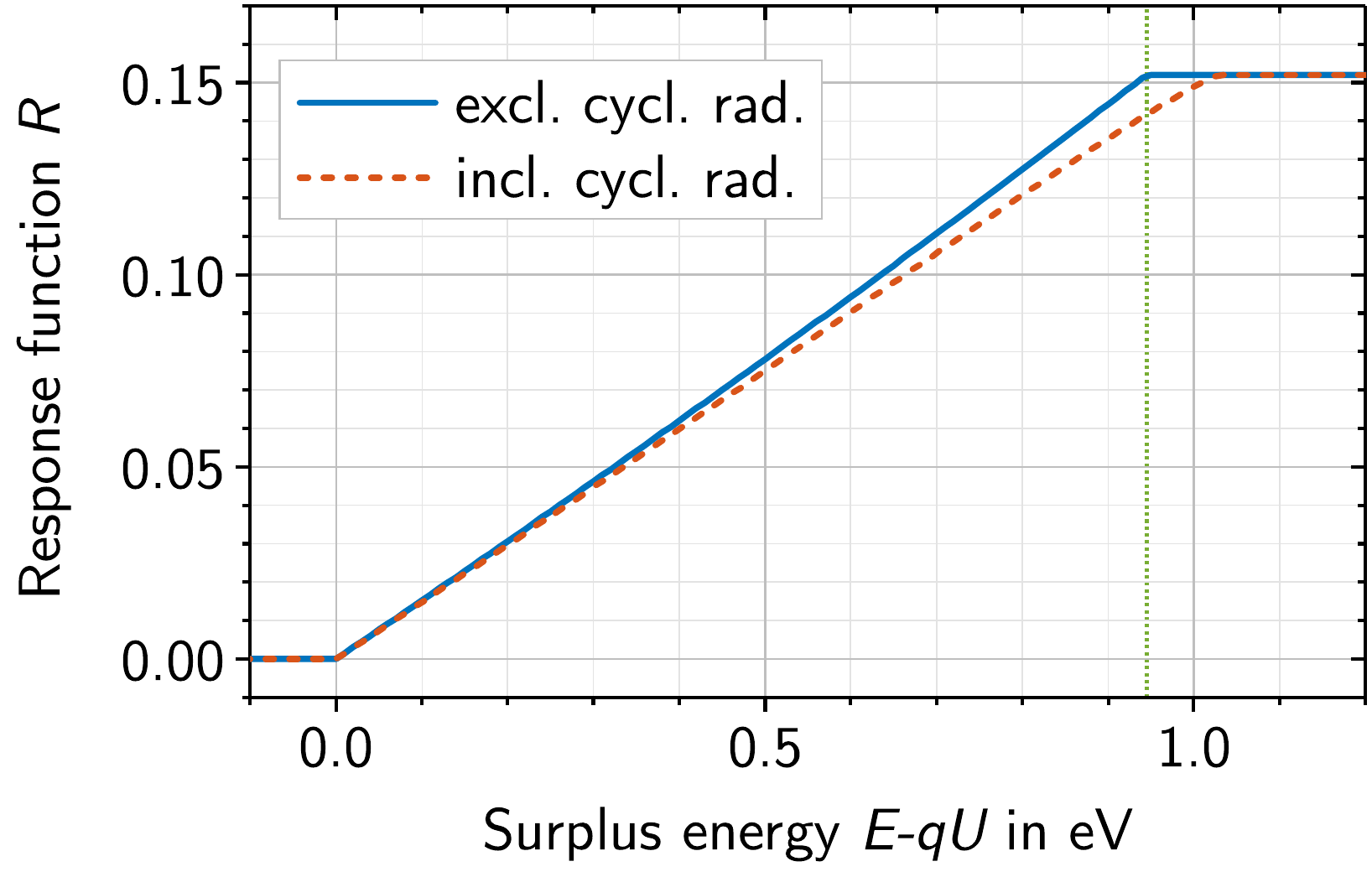}
    \caption[Cyclotron Radiation]{%
    The impact of energy losses due to cyclotron radiation on the shape of the response function near the transmission edge.}
    \label{fig:response_sync}
\end{figure}

The influence of the cyclotron energy loss on the averaged response function is shown in \cref{fig:response_sync}.

\subsection{Expected integrated spectrum signal rate}
\label{sec:integrated_spectrum}

Earlier in this section we have laid out the different contributions to the response function of the experiment, which describes the probability for \tbd electrons to arrive at the detector where they contribute to the measured integrated spectrum. The response function describes the energy analysis at the spectrometer (\cref{sec:transmission} and \cref{sec:transmission_inhomogeneity}), energy loss caused by scattering in the tritium source (\cref{sec:eloss} and \cref{sec:scattering}), and additional corrections (\cref{sec:response_noniso} and following).

Combining the response function with the description of the differential spectrum that was developed in \cref{sec:diff}, the integrated spectrum signal rate observed on a single detector pixel $j$ for a retarding potential setting $U$ can finally be expressed as
\begin{align}
        \nonumber
    \dot N_j^\text{sig}(U) &= \frac{1}{2} \, \epsilon_{\text{det},j} \cdot \sum_{i=0}^{N_L-1} \; N_{\text{T,}i} \\
      &\quad \cdot \, \int_{qU}^{\infty} \left(\frac{\td \Gamma}{\td E}\right)_\text{C,D}(m_\upnu^2, E_0) \cdot R_{i,j}(E, U) \, \td E \; .
    \label{eq:signal}
\end{align}
This expression incorporates all theoretical corrections (see \cref{eq:diff3} with subscript C) and the Doppler broadening (see \cref{eq:doppler} with subscript D) of the differential spectrum $\frac{\td \Gamma}{\td E}$ (see \cref{eq:diff2}), and the full response function which incorporates the energy loss as a result of source scattering and cyclotron radiation:
\begin{align}
        \nonumber
    R_{i,j}(E, U) &= \int_{\epsilon=0}^{\eps} \, \int_{\theta=0}^{\theta_\mathrm{max}} \, \sum_{s} \; \Tcal_{s,i,j}^\text{cycl}(E - \epsilon, \theta, U) \\
    &\quad \cdot \, P_{s,i}(\theta) \; f_s(\epsilon) \, \td\epsilon \, \sin\theta\td\theta \; .
    \label{eq:response_full}
\end{align}
The response function depends on the path traversed by the \tbd electron between its origin in source segment $(i,j)$ and the target detector pixel $j$ (see \cref{fig:seg} for the segmentation schema).
The detection efficiency $\epsilon_{\text{det},j}$ is an energy-dependent quantity, which needs to be measured for each pixel $j$. Its value is between $\approx \SI{90}{\percent}$ and $\SI{95}{\percent}$~\cite{Amsbaugh2015}.

To first order (due to nearly constant magnetic field and tritium concentration in the source), the integrated signal rate in \cref{eq:signal} depends on $\Ncal \sigma$ -- which can be accurately determined by calibration measurements with a photoelectron source -- but is independent of the longitudinal gas density profile $\rho(z)$ which cannot be measured directly (see \cite{PhDKuckert2016,arXivKuckert2018} for simulation results).

\subsection{Scan of the integrated spectrum}

A scan of the integrated \tb spectrum comprises a set of detector pixel event counts $N_j(U_k)$, observed at various retarding potential settings $U_k$ for the duration of $\Delta t_k$ each, with $k \in \{1 \ldots n_k\}$. In the following, the indices $j$ and $k$ are condensed by writing $N_{jk} = N_j(U_k)$, with $N_{jk}$ denoting the event count on a single detector pixel $j$ for a specific retarding potential setting $k$.

The observed event count $N^\text{obs}_{jk}$ is a Poisson-distributed quantity with the expectation value given by
\begin{equation}
    \text{E}[N^\text{obs}_{jk}] = \Delta t_k \cdot \left( \, \dot N_j^\text{sig}(U_k) + \dot N_j^\text{bg} \, \right) \; ,
    \label{eq:scan}
\end{equation}
where $\dot N^\text{bg}_j$ is an energy-independent background rate component (possibly with a radial dependency indicated by the index $j$).

KATRIN will be operated for a duration of 5~calendar years in order to collect 3~live years of spectrum data over multiple runs.

\subsection{Energy uncertainties}
\label{sec:escale}

At the end of this section we will briefly discuss the influence of energy uncertainties on the neutrino mass measurement.
In general, any fluctuation with variance $\sigma^2$ induces a spectrum shape deformation which -- if not considered in the analysis -- is indistinguishable to first order from a shift of the measured value of $\mnusq$ in the negative direction with $\Delta \mnusq = -2 \sigma^2$~\cite{Robertson1988}. This shift of $\Delta \mnusq$ also holds if an accounted fluctuation or distribution of true variance $\sigma^2_\mathrm{true}$ is described wrongly in the analysis by the variance $\sigma^2_\mathrm{ana} = \sigma^2_\mathrm{true} - \sigma^2$.

Different sources of fluctuations and distributions with uncertainties can be distinguished. One group comprises \tbd decay and source physics, such as molecular final states, scattering processes and the Doppler effect (all discussed in this work). Others are experimental systematics originating in the energy measurement, which have to be studied during commissioning of the setup and then incorporated into the model. An example is the distortion of the spectrometer transmission function due to retarding-voltage fluctuations~\cite{PhDKraus2016,PhDSlezak2015}.

\subsection{Impact of theoretical and experimental corrections}
\label{sec:impact}

In \cref{tab:systematics} we review and quantify the impact of theoretical corrections to the differential \tbd spectrum, discussed in \cref{sec:diff}, and of experimental corrections which have been introduced above. Many individual model components can be safely neglected, while others need to be considered more accurately, such as the radial dependence of retarding potentials (\cref{sec:transmission_inhomogeneity}), energy loss due to cyclotron radiation (\cref{sec:cyclotron}) or the Doppler effect (\cref{sec:doppler}).

\begin{table*}[tbp]
    \centering
    \begin{tabular}{l @{\hspace{3em}} r}
    \toprule[\heavyrulewidth]
    \multicolumn{2}{l}{\bfseries Source of systematic shift \hfill Systematic shift $\Delta_\text{syst}(\mnusq)$}\\[4pt]
    {\bfseries Neglected effect} or model component &  $[ \times \SI{e-5}{eV^2} ]$ \\
    \midrule[\heavyrulewidth]
    Relativistic description of $E_\text{rec}$ $^1$ & $ \num{0.03}$\\
    \midrule
    Neutrino mixing with 3 mass eigenstates (inv.\ hierarchy)& $ \num{0.04}$\\
    \midrule
    Relativistic Fermi function $F_\text{rel}(Z,E)$ $^1$ & $ \num{0.19}$\\
    \midrule[\heavyrulewidth]
    Radiative corrections ($G$) & $\num{214.10}$\\
    \midrule
    Screening correction ($S$) & $\num{-2.82}$\\
    \midrule
    Recoil, weak magnetism, \va interference corr. ($R$) & $\num{-0.12}$\\
    \midrule
    Finite nucl.\ ext.\ corr.\ ($LC$) & $< \num{0.01}$\\
    \midrule
    Recoiling Coulomb field corr.\ ($Q$) & $\num{-0.02}$\\
    \midrule
    Orbital electron exch.\ corr.\ ($I$) & $\num{-0.02}$\\
    \midrule
    Calculate $G, R, Q$ for each final state $^2$ & $\num{13.50}$\\
    \midrule[\heavyrulewidth]
    Energy loss due to elastic $\text{e}^--\text{T}_2$ scattering & $\num{-5.20}$\\
    \midrule
    Transmission function $T^\star$ (non-isotropic angular distr.) & $\num{1027.51}$\\
    \midrule
    Energy loss due to cyclotron radiation & $\num{-2939.43}$\\
    \midrule
    Radial dependence of analyzing magnetic field in $R_j(E, U)$ $^3$ & $\num{904.20}$\\
    \midrule
    Radial dependence of retarding potential in $R_j(E, U)$ & $\num{8470.47}$\\
    \midrule[\heavyrulewidth]
    Doppler effect (thermal and bulk velocity neglected) & $\num{-1554.46}$\\
    \midrule
    Doppler effect (only bulk gas velocity neglected) & $\num{117.81}$\\
    \midrule
    Doppler effect (only approximated by smearing the FSD) & $\num{101.41}$\\
    \bottomrule[\heavyrulewidth]
    \end{tabular}
    \caption[Corrections and systematic uncertainties]{Impact of individual theoretical and experimental model corrections on the measured squared neutrino mass $\mnusq$, if neglected or approximated. The analysis energy window is restricted to $[E_0 - \SI{30}{eV}; E_0 + \SI{5}{eV}]$. For $\mnu$ a true value of \SI{200}{meV} is assumed.\\
	$^1$ Instead of using the non-relativistic variant.\\
	$^2$ Instead of pulling these effects outside the FSD summation in \cref{eq:diff3}.\\
	$^3$ With a central analyzing magnetic field $B_\text{ana} = \SI{3.6E-4}{T}$.
    }
    \label{tab:systematics}
\end{table*}

\section{Measurement of the neutrino mass}
\label{sec:measurement}

Having compiled a complete description of the theoretical \tbd decay spectrum and the response function of KATRIN into a parameterizable model, we will now outline the statistical terms and methods required for actual neutrino mass measurements.
In the next (\crefrange{sec:params}{sec:conf}) we review the process of parameter inference (model fitting) and the construction of confidence intervals in the case of a KATRIN neutrino mass analysis, and we explain the relation between observed data, fit parameters and their uncertainties. After introducing Frequentist methods of inferring $\mnusq$ we give an example of a Bayesian approach in \cref{sec:bayesian}. We briefly list statistical and systematic uncertainty contributors for KATRIN in \cref{sec:statsys} and in that context discuss the relevance of the choice of the energy analysis interval in \cref{sec:energyinterval} and the distribution of accounted measuring time among that interval in \cref{sec:mtd}. In \cref{sec:negmnusq} we give an explanation of negative $\mnusq$ estimates and provide a non-physical extension of the \tbd decay spectrum model.

\subsection{Parameter inference}
\label{sec:params}

The statistical technique for analyzing \tbd decay spectrum data is well established. By comparing the observed number of counts $N^\text{obs}_{jk}$ on each pixel $j$ for each experimental setting $k$ with the prediction from the spectrum and response model $N_{jk}(U_k, \mnusq, E_0, \dots)$ (see \cref{eq:signal} and \labelcref{eq:scan}), $\mnusq$ and other unknown model parameters can be inferred. In the case of a KATRIN-like neutrino mass measurement, a continuous model that depends on $\mnusq$ is fit to unbinned spectral shape data. The method of least squares is most commonly applied.

The probability to have an observed outcome $\boldsymbol{N^\text{obs}} = \left(N^\text{obs}_{1,1} \ldots N^\text{obs}_{n_j,n_k}\right)$, given the predicted number of counts $\boldsymbol{N^\text{pre}}(\boldsymbol{\theta})$ defined by a set of model parameters $\boldsymbol{\theta} = (m_\upnu^2, E_0, \dots)$, is the likelihood function
\begin{equation}
    \label{eq:likelihood}
    L(\boldsymbol{\theta} | \boldsymbol{N^\text{obs}}) = \prod_{jk} \; \text{Poisson}\left( N^\text{obs}_{jk} | N^\text{pre}_{jk}(\boldsymbol{\theta}) \right) \; .
\end{equation}
A set of parameter point estimates $\boldsymbol{\hat\theta}$ is obtained by maximizing the likelihood $L$.  Equivalently, a minimization of the negative log-likelihood $- \ln L$ can be performed, which is often more practical numerically.

If the number of observed events $N_{jk}^\text{obs}$ is large enough ($\gtrsim 25$), so that the Poisson distribution can be approximated by a Gaussian, that expression is approximately a $\chi^2$ function:
\begin{equation}
    \label{eq:chi2}
    - 2 \ln L \approx \chi^2 = \sum_{jk} \left(\frac{ N^\text{obs}_{jk} - N^\text{pre}_{jk}(\boldsymbol{\theta}) }{\sigma_{jk}}\right)^2 \; .
\end{equation}
In case of $\sigma_{jk} = \sqrt{N_{jk}^\text{pre}}$, the above $\chi^2$ equals the Pearson's chi-square statistic~\cite{bib:chisquare}.

Our \emph{parameter of interest} is $\mnusq$, which distorts the spectrum shape close to the endpoint. Because the fitted \tbd spectrum shape essentially only depends on $\mnusq$, with $\chi^2$ being approximately parabolic in $\mnusq$, it is the preferred fit parameter over $\mnu$~\cite{Holzschuh1992}.

Other model parameters are \emph{nuisance} parameters. In KATRIN-like experiments typically three such quantities are treated as free fit parameters:
\begin{itemize}
  \item
    The tritium endpoint energy $E_0$, the maximum electron energy assuming a vanishing neutrino mass, has to be estimated from the data, due to uncertainties in the measured $\chem{T^+}$/$\chem{^3He^+}$ mass difference~\cite{bib:myers} and in the experimental energy scale.

  \item
    The signal amplitude $A_\text{sig}$, a multiplicative factor close to \num{1}, is applied to the predicted signal rate%
        \footnote{Deviations from unity arise mainly from  incomplete knowledge of the tritium column density and the detector efficiency (see \cref{eq:signal}).}%
    $\dot N_j^\text{sig}$ to correct for any energy-independent model uncertainty. $E_0$ and $A_\text{sig}$ are estimated from the slope of the spectrum at lower energies of the analysis interval ($\approx \SIrange{30}{40}{eV}$ below the endpoint), where the absolute signal rate is highest.

  \item
    The background rate amplitude $A_\text{bg}$ is another normalization factor, which is applied to the background model component $\dot N_j^\text{bg}$. It is estimated using the data from retarding potentials above the tritium endpoint, where no signal is expected. Note that we assume a constant background rate without retarding potential dependence in the energy interval near the tritium endpoint. However, such an energy dependence could be incorporated into the model using additional data above the endpoint.

\end{itemize}

Considering only the aforementioned four model parameters, the predicted number of electrons on a detector pixel $j$ for a retarding potential setting $k$ in a counting period $\Delta t_k$ is given by
\begin{align}
        \nonumber
    &N^\text{pre}_{jk}(\mnusq, E_0, A_\text{sig}, A_\text{bg}) =\\
      &\qquad \Delta t_k \cdot \left( \, A_\text{sig} \cdot \dot N_j^\text{sig}(U_k, m_\upnu^2, E_0) \, + \, A_\text{bg} \cdot \dot N_j^\text{bg} \, \right) \; .
\end{align}
A point estimate for this set of parameters, obtained from maximizing the likelihood (or minimizing $\chi^2$) is denoted in the following as $(\hmnusq, \widehat{E_0}, \widehat{A_\text{sig}}, \widehat{A_\text{bg}})$.

Depending on the method of treating systematic uncertainties, the number of free (or constrained) model parameters can be higher.

\subsection{Confidence intervals}
\label{sec:conf}

Due to the stochastic nature of the observed data, a single parameter point estimate by itself cannot relate to the unknown true value of a parameter. In parameter inference, a confidence interval defines an interval of parameter values that contain the true value of the parameter to a certain proportion (confidence level), assuming an infinite number of independent experiments. Various methods of constructing such intervals exist.

\begin{figure}[ht]
    \centering
    \includegraphics[width=\figurewidthonecol]{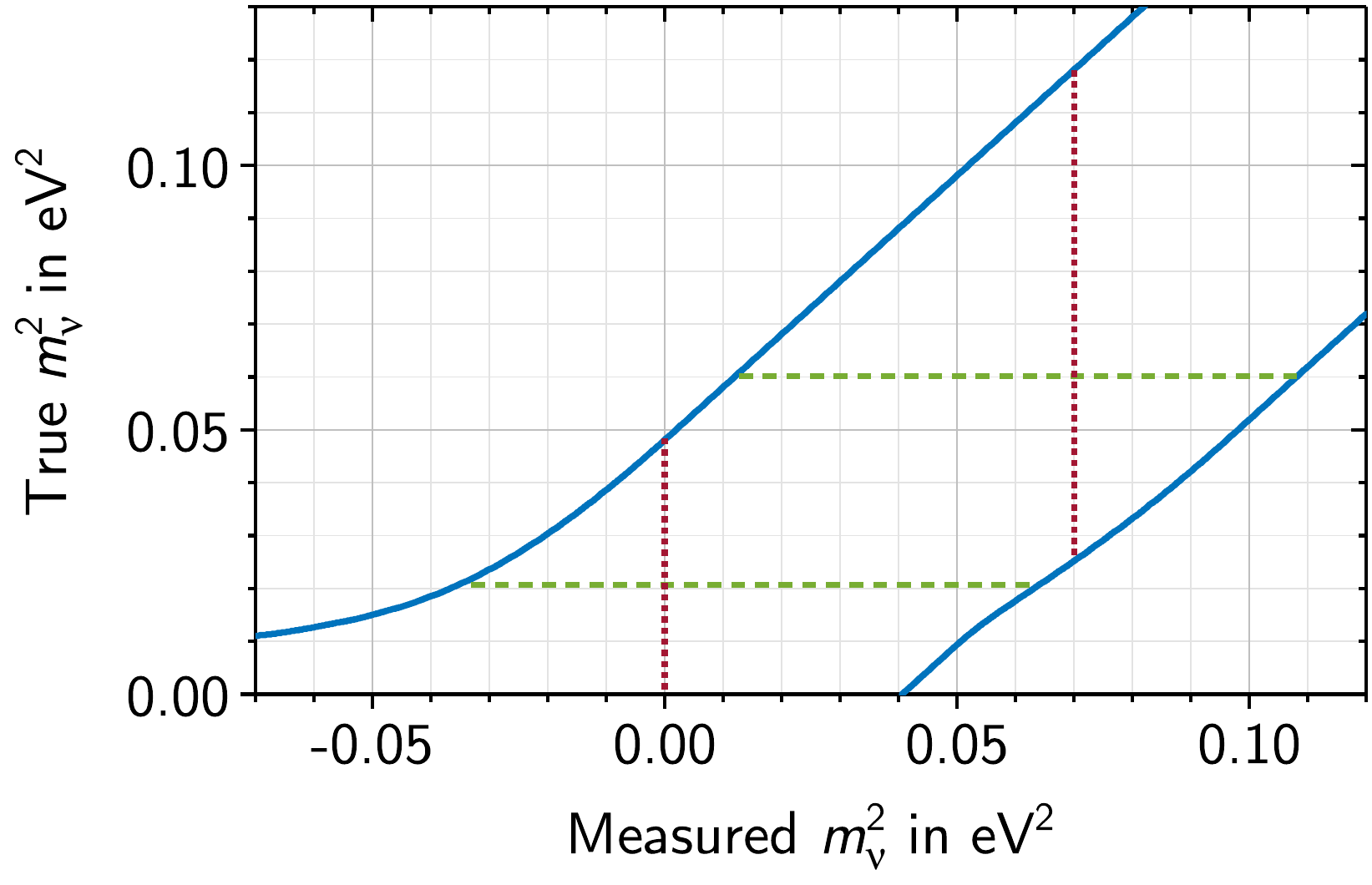}
    \caption[Unified Confidence Belt]{%
    Frequentist confidence belt (\SI{95}{\percentcl}) constructed according to the unified approach by Feldman and Cousins~\cite{bib:feldman}. In this example, the horizontal ranges (green dashed lines) are constructed by choosing \SI{95}{\percent} of the $\mnusq$ estimates from an ensemble test with fixed true $\mnusq$, following the ordering principle. These horizontal ranges define the edges of the confidence belt (blue solid lines).
    The subsequent result of an actual neutrino mass measurement (x-axis, indicated by red dotted lines) is used to select the vertical intersections with the confidence belt to determine the reporting of an upper limit (e.g.\ in case of $\mnusq = \SI{0}{eV^2}$) or a two-sided confidence interval (e.g.\ in case of $\mnusq = \SI{0.07}{eV^2}$).
    }
    \label{fig:unified}
\end{figure}

Using the Neyman construction~\cite{bib:neyman} (a Frequentist method), ensembles of pseudo-experiments are sampled for a range of true values of $\mnusq$, leading to the construction of a confidence belt (see \cref{fig:unified}). Incorporating an ordering principle proposed by Feldman and Cousins~\cite{bib:feldman,bib:karbach}, empty confidence intervals for non-physical estimates of $\mnusq$ can be avoided, while ensuring correct Frequentist coverage.

When parameter point estimates are constructed following the maximum likelihood ordering principle, the profile likelihood ratio~\cite{bib:rolke} can be used to estimate their uncertainties. With this method the $\SI{1}{\sigma}$ uncertainty of a parameter estimate is identified by those parameter values where the likelihood has decreased to half its maximum value, while profiling (maximizing) with respect to any involved nuisance parameter.
Equivalently, a chi-square curve can be scanned for parameter values with $\Delta \chi^2 = 1$, again profiling over nuisance parameters.

\begin{figure*}[tbp!]
    \centering
    \includegraphics[width=0.95\figurewidthtwocols]{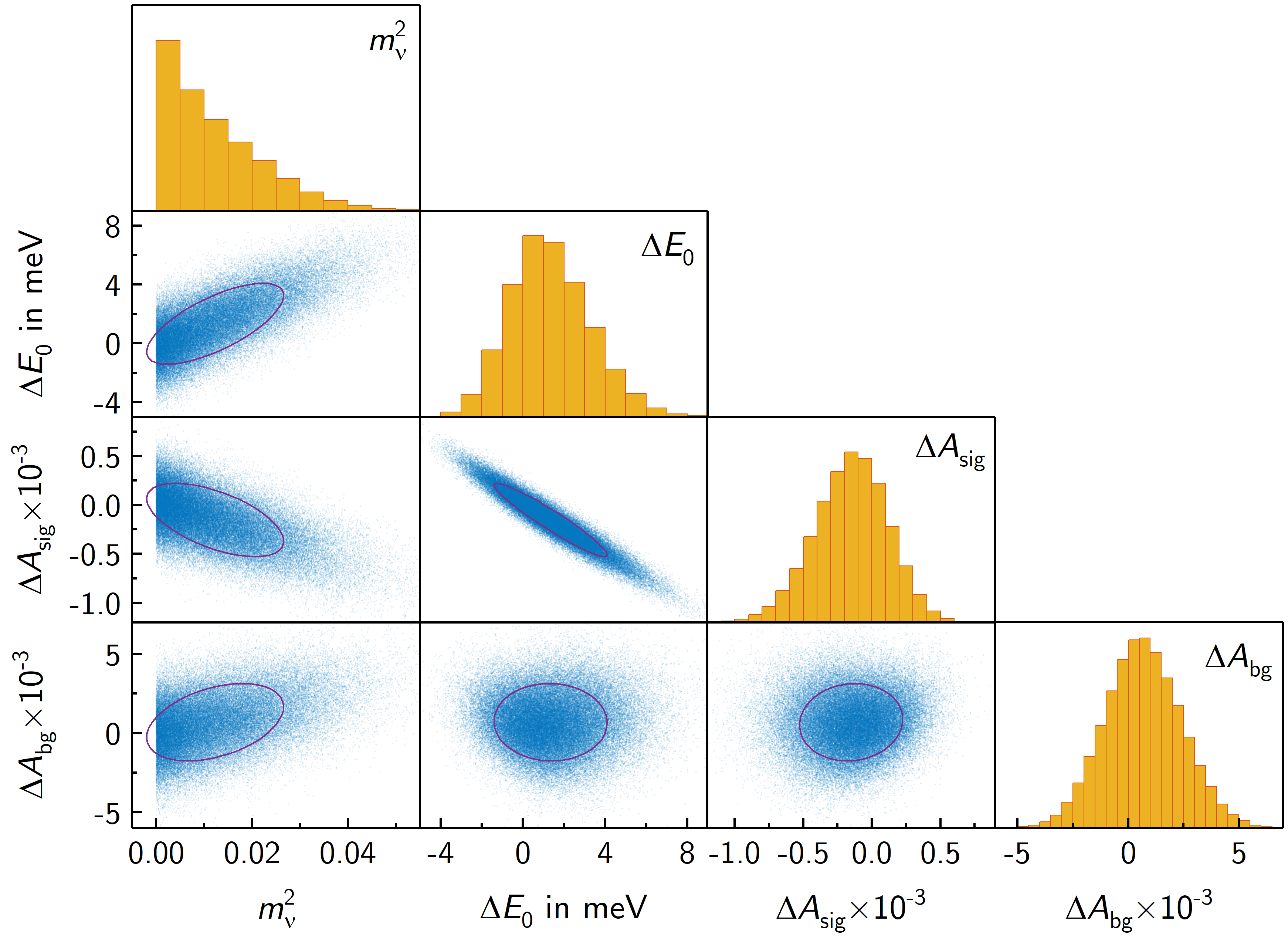}
    \caption[Four parameter scatter plot]{%
    Scatter plots for pair-wise parameter combinations ($\mnusq$, $E_0$, $A_\text{sig}$, $A_\text{bg}$) and their respective marginalized posterior distributions as the diagonal elements for \SI{3}{years} of live measurement time. The solid contours indicate \SI{95}{\percentcl} regions.
    Instead of randomized data, the likelihood sampled in this MCMC example was formulated based on a null hypothesis with fiducial input values $\mnusq = \SI{0}{eV^2}$, $E_0 = \SI{18575}{eV}$, $A_\text{sig} = \num{1.0}$, $A_\text{bg}\cdot\dot N^\text{bg} = \SI{10}{mcps}$. Flat priors were used with $\mnusq \ge \SI{0}{eV}$.
    }
\label{fig:scatter}
\end{figure*}

\subsection{Bayesian statistics}
\label{sec:bayesian}

Bayesian inference is typically based on the posterior PDF (probability density function) of a parameter of interest. Using Bayes' theorem, the posterior distribution $p(\boldsymbol{\theta})$ of a set of parameters $\boldsymbol{\theta}$ is given by the likelihood $L(\boldsymbol{\theta})$ and a prior probability $\pi(\boldsymbol{\theta})$:
\begin{equation}
    p(\boldsymbol{\theta}) \propto L(\boldsymbol{\theta}) \cdot \pi(\boldsymbol{\theta}) \; .
\end{equation}
In contrast to Frequentist approaches, which make a statement about the repeatability of an experiment, Bayesian statistics inevitably introduce the concepts of probability, belief and credibility. The prior probability $\pi(\boldsymbol{\theta})$ has to be chosen by the analyst, based on prior belief. In the case of $\mnusq$, an objective option is the flat uniform prior (possibly zero for $\mnusq < \SI{0}{eV^2}$), or a normalizable Gaussian distribution that reflects the results from previous measurements.

Fortunately, KATRIN's $\mnusq$ posterior PDF is rather insensitive to the choice of prior on $\mnusq$. Assuming, for instance, a true value of $\mnusq = \SI{0}{eV^2}$, a Gaussian prior with mean $\mu_\pi = \SI{0}{eV^2}$ and $\sigma_\pi = \SI{1}{eV^2}$ (or a value on the order of the Mainz or Troitsk upper limits) will be outweighed by the KATRIN likelihood function. It will thus have no significant effect on the derived Bayesian upper limit compared to a prior that is flat in $\mnusq$. This underlines the improved sensitivity of the experiment.

The posterior distributions can be obtained practically with Markov-chain Monte Carlo (MCMC) methods~\cite{bib:robert}. With proper adjustments, this class of algorithms is capable of efficiently traversing high-dimensional parameter spaces and sampling from posterior probability distributions of an unknown quantity such as $\mnusq$. From these distributions, any choice of credibility interval $[\theta_1, \theta_2]$, with $P = \int_{\theta_1}^{\theta_2} p(\boldsymbol{\theta}) \td \theta$ being the confidence level, can be constructed.

When considering the distribution of only a subspace of all parameters, one speaks of a marginal posterior distribution.  To determine the one-dimensional posterior distribution of  $\mnusq$, the four-dimensional posterior distribution of $(\mnusq, E_0, A_\text{sig}, A_\text{bg})$ is marginalized over the three nuisance parameters.

\Cref{fig:scatter} shows the result of a MCMC sampling of the posterior distribution that uses the basic Metropolis-Hastings~\cite{bib:metropolis} algorithm. The underlying model is based on \cref{eq:likelihood} with its standard four model parameters $(\mnusq, E_0, A_\text{sig}, A_\text{bg})$, using flat priors and the constraint $\mnusq \ge \SI{0}{eV^2}$. In this representation, the correlations between these parameters can be assessed easily. The correlation matrix of this particular example evaluates to:

{
\centering
\vspace{6pt}
\renewcommand{\arraystretch}{1.3}
\begin{tabular}{c | c c c c}
	& $\mnusq$      & $E_0$         & $A_\text{sig}$ & $A_\text{bg}$ \\
	\hline
	$\mnusq$        & \num{1}       &               &               &           \\
	$E_0$           & \num{0.698}   & \num{1}       &               &           \\
	$A_\text{sig}$  & \num{-0.581}  & \num{-0.953}  & \num{1}       &           \\
	$A_\text{bg}$   & \num{0.396}   & \num{-0.022}  & \num{0.077}   & \num{1}   \\
\end{tabular}
\par
\vspace{6pt}
}

\begin{figure}[ht]
    \centering
    \includegraphics[width=\figurewidthonecol]{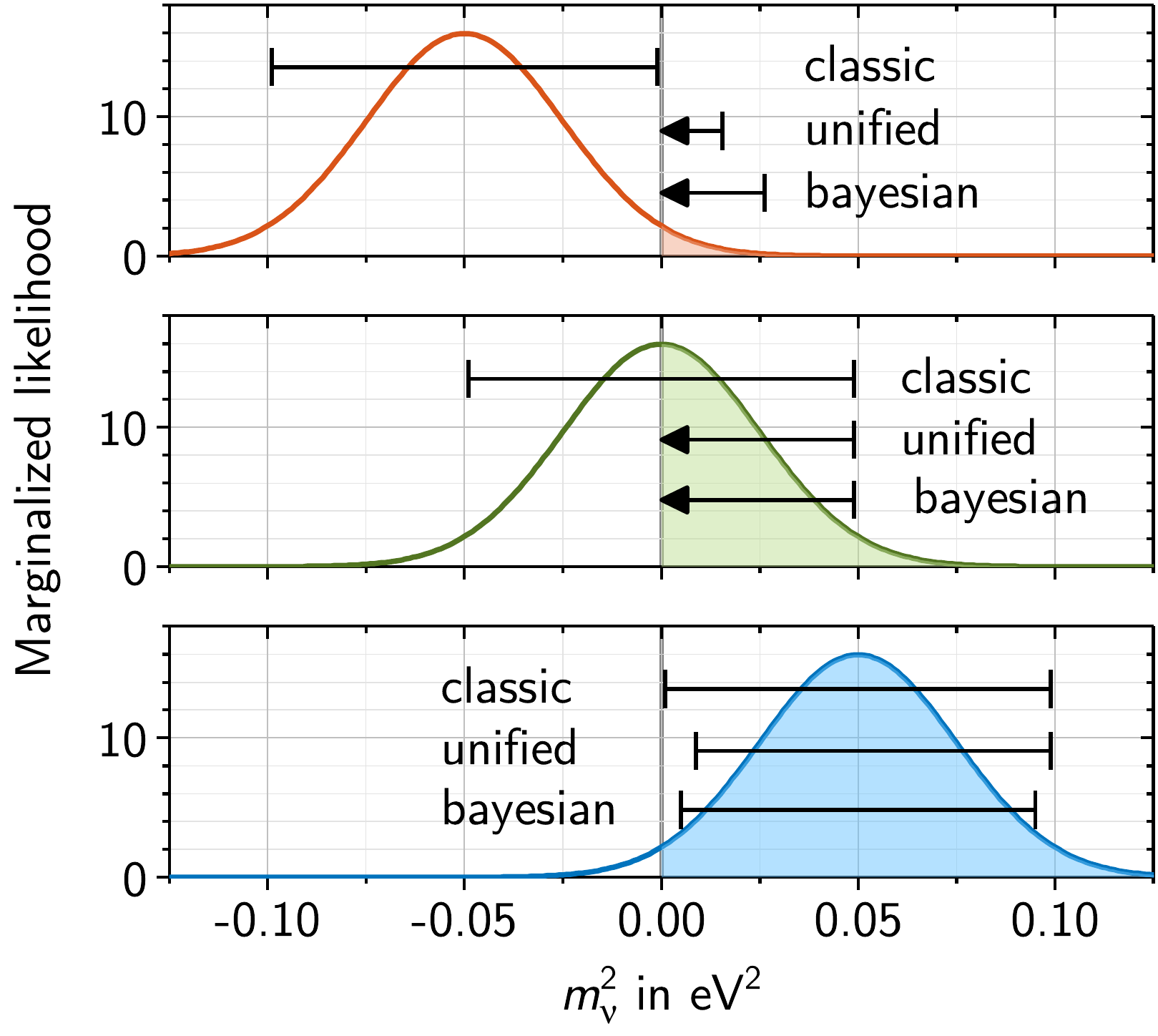}
    \caption[Confidence intervals]{%
    Marginalized likelihood functions for various estimates of $\mnusq$ from selected sets of simulated data. Top panel: $\hmnusq = \SI{-0.05}{eV^2}$. Middle panel: $\hmnusq = \SI{0.0}{eV^2}$. Bottom panel: $\hmnusq = \SI{0.05}{eV^2}$ ($\hmnu = \SI{225}{meV}$). The horizontal bars indicate \SI{95}{\percentcl} Frequentist central confidence intervals (\emph{classic}), Feldman and Cousins (\emph{unified}), and Bayesian credibility intervals (\emph{Bayesian}) with a flat prior for $\mnusq \geq \SI{0}{eV^2}$.
    }
    \label{fig:cmp}
\end{figure}

A comparison of Bayesian and Frequentist confidence intervals for various estimates of $\mnusq$ is given in \cref{fig:cmp}. For positive estimates, the different methods yield similar results.

\subsection{Statistical and systematic uncertainties}
\label{sec:statsys}

Traditionally, the statistical uncertainty $\sigma_\text{stat}(\mnusq)$ is identified with the spread of an $\mnusq$ estimate caused by the randomness of the observed data (spectrum count rates $N^\text{obs}_{k}$), and usually decreases when data are taken (as $1/\sqrt{N_k}$ or $1/\sqrt{\Delta t_k}$).
A systematic uncertainty $\sigma_\text{syst}(\mnusq)$, by contrast, represents an uncertainty in the $\mnusq$ estimate due to an uncertainty in the spectrum or response model which does not scale with the amount of data taken in general.

Providing a comprehensive review of all systematics of KATRIN  -- some of which are not adequately quantifiable until final commissioning and characterization of the experimental apparatus -- is beyond the scope of this article.
Among the major systematic contributors are the final state distribution (\cref{sec:fsd}), the shape of the energy loss function and the inelastic scattering cross section (\cref{sec:eloss}), the source-gas column density (\cref{sec:source}), and high-voltage fluctuations (\cref{sec:escale}).

\begin{figure}[ht]
    \centering
    \includegraphics[width=\figurewidthonecol]{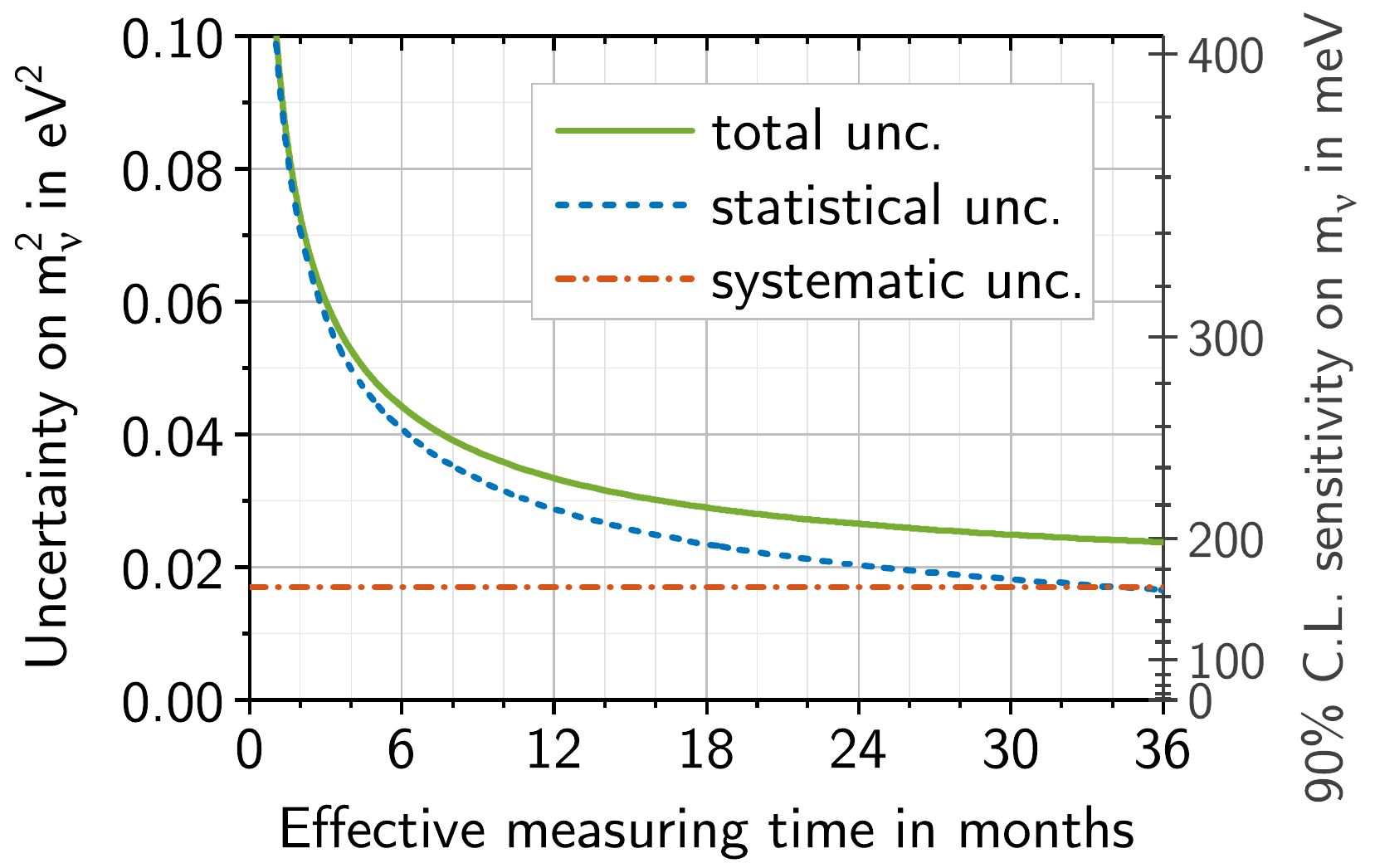}
    \caption[Sensitivity over time]{%
    Statistical, systematic and total $\SI{1}{\sigma}$ uncertainty of $\mnusq$ on the left vertical axis, and \SI{90}{\percentcl} sensitivities of $\mnu$ on the right vertical axis, plotted over the effective measuring time. Thirty-six live months (3 live years) correspond to 5 calendar years of KATRIN operation.
    }
    \label{fig:sensitivity}
\end{figure}

The total systematics budget of KATRIN is conservatively evaluated to a maximum value of $\sigma_\text{syst}(\mnusq) \approx \SI{0.017}{eV^2}$~\cite{KATRIN2005}. Accordingly, KATRIN's setup and configuration are chosen in such a way that the statistical uncertainty, after an envisaged data-taking period of five calendar years, reaches $\sigma_\text{stat}(\mnusq) \approx \sigma_\text{syst}(\mnusq) \approx \SI{0.017}{eV^2}$, as depicted in \cref{fig:sensitivity}.
These values are commonly translated into a \SI{90}{\percentcl} sensitivity of
\begin{equation}
    S(\mnu) = \sqrt{ 1.645 \cdot \sigma_\text{tot}(\mnusq) } \approx \SI{200}{meV}
\end{equation}
with the total uncertainty on $\mnusq$
\begin{equation}
    \sigma_\text{tot}(\mnusq) = \sqrt{ \sigma_\text{stat}^2(\mnusq) \, + \, \sigma_\text{syst}^2(\mnusq) } \; .
\end{equation}

\subsection{Choice of the analysis energy interval}
\label{sec:energyinterval}

\begin{figure}[h!]
	\centering
	\includegraphics[width=\figurewidthonecol]{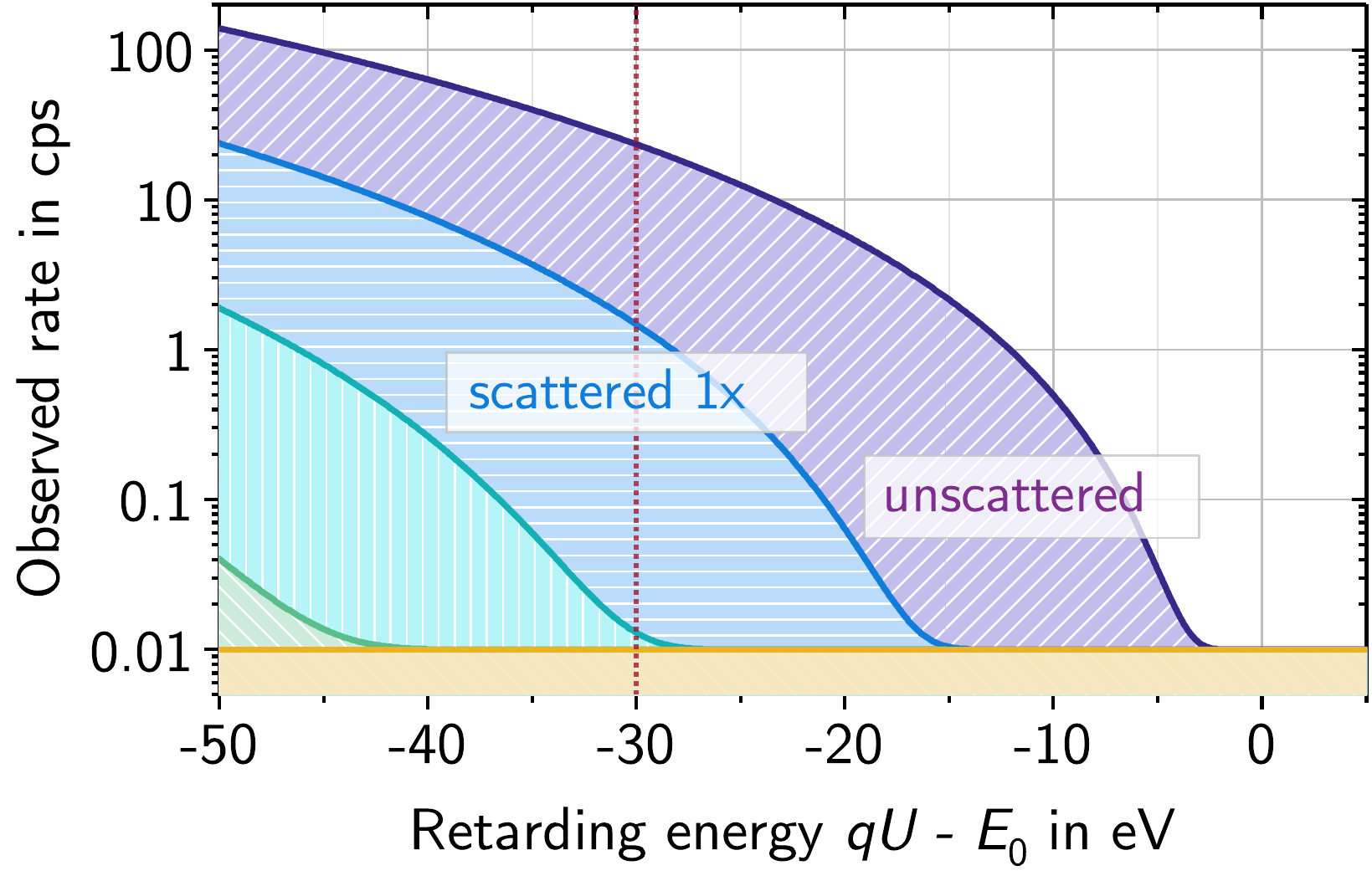}
	\caption[Population of scattered electrons]{%
		The expected \tbd spectrum rate with different shaded areas depicting the fraction of scattered and unscattered electrons. The lower baseline comprises the \SI{10}{mcps} energy-independent background component. Starting from the right, the shaded areas comprise signal \tbd electrons that are unscattered, scattered once, twice, and thrice.
	}
	\label{fig:scatprob}
\end{figure}

The optimal choice of the lower spectrum energy threshold for analysis is primarily determined by the ratio of the statistical and systematic uncertainties. Neither one should dominate. With the differential spectrum rising quadratically as the filter energy $qU$ is lowered (for $E_0-E \gg \mnu$), the statistical uncertainty on the observed number of signal electrons $\sigma_\text{stat}\left(N_{j}^\text{sig}(qU_k)\right)$ decreases. On the other hand, systematic uncertainties due to energy-loss processes or electronic excitations of the daughter molecule increase at lower energies. Assuming the design operational configuration of KATRIN (see \cref{tab:design}), a lower threshold of $E_0-\SI{30}{eV}$ will lead to the desired alignment of statistical and total systematic uncertainties ($\sigma_\text{stat}(\mnusq) \approx \sigma_\text{syst}(\mnusq)$ ). As shown in \cref{fig:scatprob}, the spectrum in this energy range is mainly populated with electrons that have scattered off the source gas at most once.

\subsection{Measuring time distribution}
\label{sec:mtd}

\begin{figure*}[tbp!]
    \centering
    \includegraphics[width=\figurewidthtwocols]{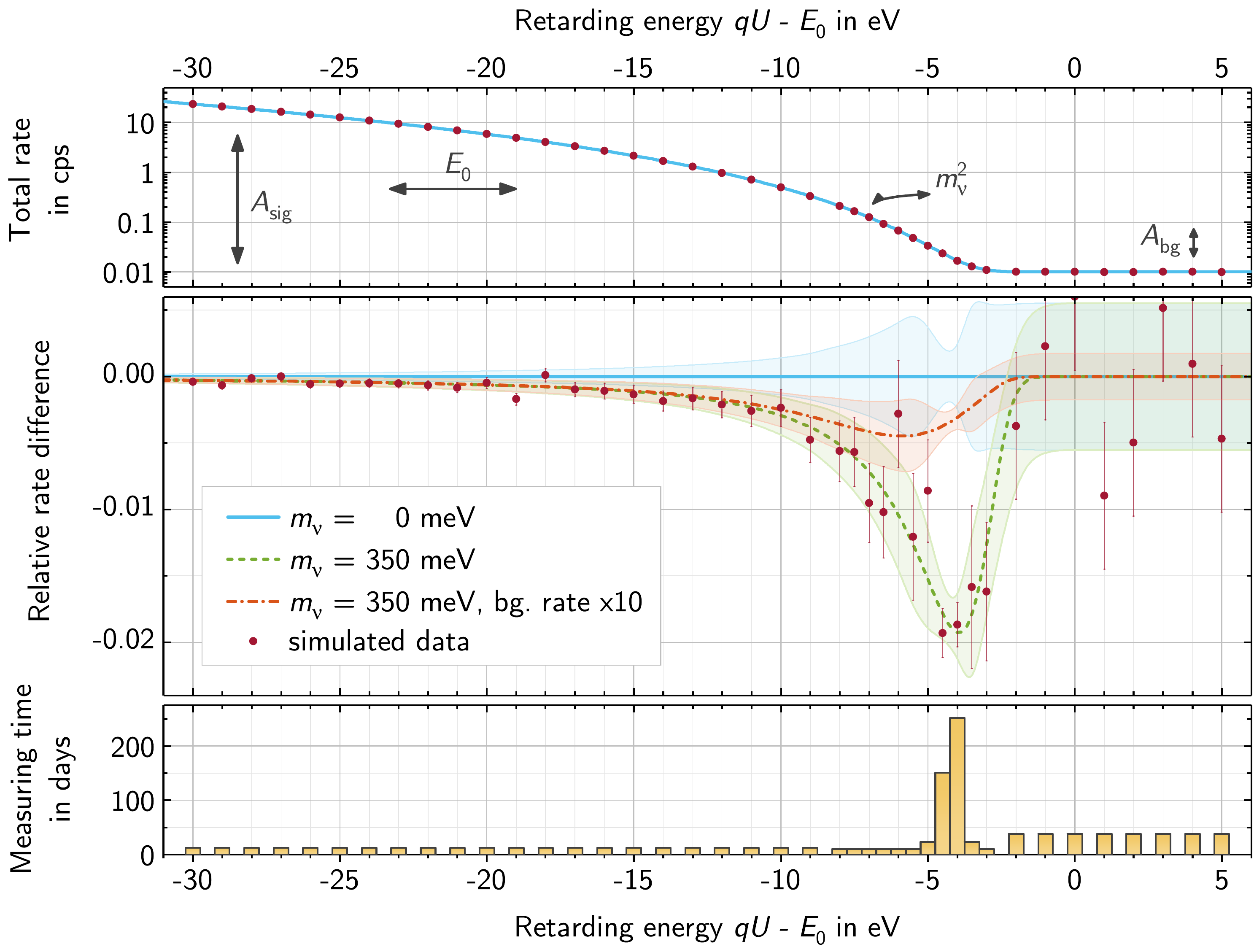}
    \caption[Relative spectrum rates and measuring time]{%
    An illustration of a hypothetical neutrino mass signal, using toy data simulated for $\mnu = \SI{350}{meV}$ (red points + stat.\ error bars), compared against the theoretical model expectations for $\mnu = \SI{0}{meV}$ (blue solid line), $\mnu = \SI{350}{meV}$ (green dashed line) at nominal background of $R_\text{bg} = \SI{10}{mcps}$, and $\mnu = \SI{350}{meV}$ at elevated background $R_\text{bg} = \SI{100}{mcps}$ (orange dash-dotted line).

    Top panel: The absolute rate $\sum_{j}\dot N_{jk}(U_k) = N_{k}(U_k)$ is plotted against the retarding energy $qU_k$ relative to the endpoint energy $E_0$.

    Middle panel: The relative rate difference near the endpoint energy. Under the nominal background conditions, the largest deficit in rate due to a non-zero neutrino mass is expected to be about \SI{4}{eV} below the endpoint, where the signal-to-background ratio is $\approx 1$. For the scenario of a higher background rate, this point of maximal distortion is shifted to lower energies. The shaded bands indicate the statistical uncertainties.

    Bottom panel: The measuring time $\Delta t_k$ attributed to each retarding potential setting $U_k$. The Poisson uncertainty of the generated toy rates $\dot N_k$ is directly related to the measuring time through $\sigma(\dot N_k) = \sqrt{\dot N_k \, / \, \Delta t_k}$.}
    \label{fig:schedule}
\end{figure*}

The distribution of measuring time $\Delta t_k$ over a range of retarding potentials is of particular importance. Because the statistical uncertainties of the observed Poissonian rates are given by
\begin{equation}
    \sigma(\dot N) = \sqrt{N} / \, t = \sqrt{\dot N / \, t} \; ,
\end{equation}
more measuring time should be allocated to those regions of the spectrum that are most effective for estimating the parameters of interest and the correlated nuisance parameters.

\Cref{fig:schedule} illustrates the relative spectrum rates with a measuring time distribution in the energy interval of $[E_0-\SI{30}{eV}, E_0+\SI{5}{eV}]$.
In the case of $\hmnusq$, sufficient measuring time must be spent on the region slightly below the endpoint, where the spectral distortion due to a non-zero $\mnu$ is most prominent. This is also the region with a signal-to-background ratio between $2:1$ and $1:1$. Accordingly, for scenarios of elevated background, this feature of the measuring time distribution must be adapted and shifted to slightly lower energies.

\begin{figure}[h!]
	\centering
	\includegraphics[width=\figurewidthonecol]{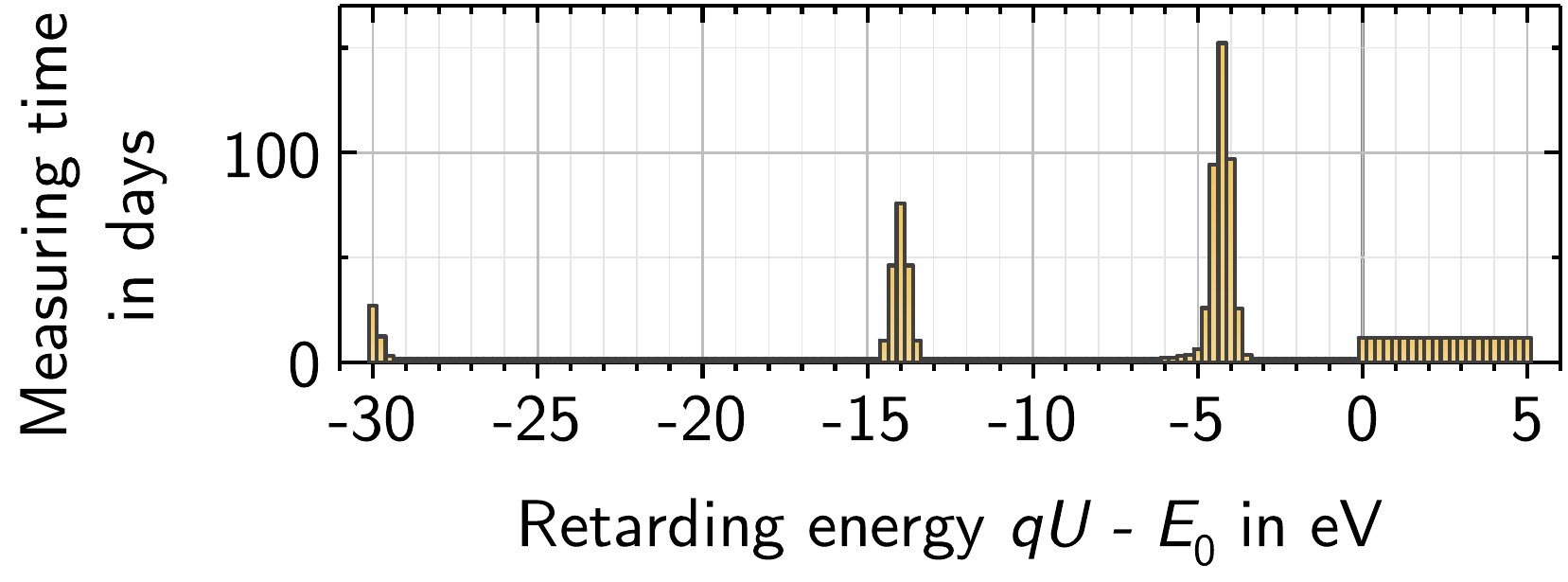}
	\caption[Optimized measuring time distribution]{%
		The measuring time $\Delta t_k$ attributed to various retarding potential settings $U_k$ in a more sparse, statistically optimized distribution.
	}
	\label{fig:mtd_focused}
\end{figure}

The measuring time distribution can be further optimized to provide even better statistical leverage on the model parameters fit to the spectrum shape (see \cref{sec:params}), reducing the statistical uncertainty $\sigma_\text{stat}^\text{opti}\left(\mnusq\right) < \SI{0.015}{eV^2}$ for nominal experimental conditions~\cite{PhDKleesiek2014}.
An example is shown in \cref{fig:mtd_focused}, which describes a rather sparse measuring time distribution with only four features, covering distinct retarding energy regions $qU$.
The peak at the lower end of the analysis energy interval ($\approx\SI{-30}{eV}$) is best suited to measure $E_0$ and $A_\text{sig}$ due to the higher absolute spectrum rates. At $qU-E_0\approx\SI{-14.0}{eV}$ the correlation between $E_0$ and $A_\text{sig}$ is broken. $\mnusq$ is measured through the \tb spectrum shape distortion around $qU-E_0\approx\SI{-4.5}{eV}$, where about one third of the overall measuring time is invested. $A_\text{bg}$ is measured using data beyond the endpoint energy $E_0$, where no \tbd decay signal is expected. Note that all four of these parameters are correlated, so the measuring time cannot be shifted arbitrarily between these four regions of retarding energy.

This more focused model allows a lower statistical uncertainty of the measured $\mnusq$, however, it bears a higher risk of overseeing unexpected spectrum shape distortions in the neglected regions of the \tbd decay spectrum.
To safeguard against such spectral deviations from the model and against unexpected systematics, a more uniform distribution, such as the one first shown in \cref{fig:schedule}, seems more appropriate, at least for the initial data-taking period.

\subsection{Negative $\mnusq$ estimates}
\label{sec:negmnusq}

The true value of $\mnusq$ is expected to be very close to $\mnusq = \SI{0}{eV^2}$~\cite{Drexlin2013}. Assuming non-tachyonic neutrinos, the physical lower limit of the effective neutrino mass is given by the neutrino mass eigenstate splittings, measured by neutrino oscillation experiments~\cite{Robertson1988}.

\begin{figure}[h!]
    \centering
    \includegraphics[width=\figurewidthonecol]{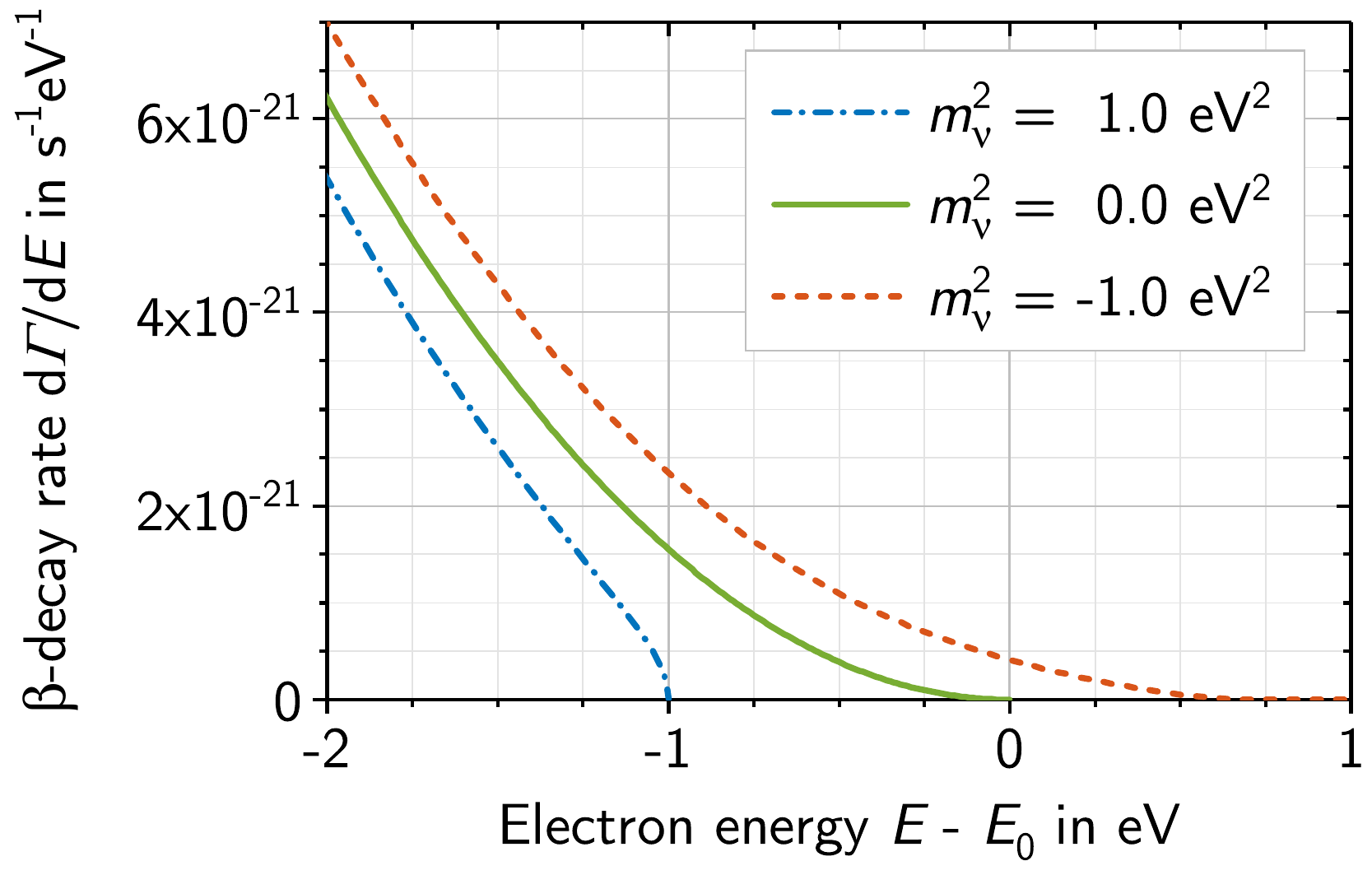}
    \caption[Non-physical spectrum extrapolation]{%
    Extrapolation of the differential \tb spectrum model for different values of the measured neutrino mass squared, including an nonphysical value of $\hmnusq = \SI{-1}{eV^2}$ (dashed red line).
    }
    \label{fig:nonphys}
\end{figure}

In order to adequately follow statistical fluctuations of the data in a $\chi^2$ parameter fit, it is necessary to allow the estimator $\hmnusq$ to take values beyond the physical limit. This is achieved by using a non-physical continuous extrapolation of the spectrum (see \cref{fig:nonphys}), which modifies the differential \tb spectrum in \cref{eq:diff3} by
\begin{equation}
    \epsilon_f \, \sqrt{\epsilon_f^2-\mnusq} \; \longrightarrow \; \biggl( \epsilon_f + \mu \, \upe^{ -\epsilon_f / \mu - 1 } \biggr) \; \sqrt{\epsilon_f^2-\mnusq}
\end{equation}
with $\mu = k \, \sqrt{ - \mnusq}$ for $\mnusq < 0$ and $\mu = 0$ for $\mnusq > 0$.
The factor $k \approx \num{0.72}$ is adjusted based on numerical calculations to make the $\chi^2(\mnusq)$ function (and negative log-likelihood respectively) symmetric around its minimum. This construction ensures a symmetric and continuous distribution of $\hmnusq$ estimates, approximating a standard normal distribution even close to the physical boundary.
A similar extrapolation scheme was used in the analysis of the Mainz and Troitsk neutrino mass experiments~\cite{Kraus2005,Aseev2011}.

The interpretation of negative $\hmnusq$ estimates and the construction of physical confidence intervals for $\mnu$ are handled differently in Frequentist and Bayesian statistics. The unified approach~\cite{bib:feldman} is aimed at constructing intervals for nonphysical parameter estimates with correct Frequentist coverage (see also \cref{fig:unified}). In a Bayesian framework the prior for $\mnusq$ is typically set to 0 for values of $\mnusq < \SI{0}{eV^2}$, making the above extrapolation redundant.

\section{Conclusion}
\label{sec:conclusion}

Using \tb spectroscopy, the KATRIN experiment aims to probe the absolute neutrino mass scale with an unprecedented sub-eV sensitivity. Both the statistical and systematic uncertainties of the model parameter of interest, the squared electron neutrino mass $\mnusq$, are required to be on the order of $\mathcal{O}(\SI{0.01}{eV^2})$. This demands a solid understanding and consistent implementation of the theoretical \tbd decay spectrum model and the experimental response function.

With this work, an effort was made to summarize the \tb spectrum calculation with all known theoretical corrections relevant for spectroscopy in the endpoint region. Furthermore, a response function model of the KATRIN experiment was outlined, including its dependencies on source-gas dynamics and the spectrometer electromagnetic configuration. Finally, the statistical methods applicable to the intended measurement were investigated and concrete examples of their application to the KATRIN neutrino mass measurement were given.

In \cref{sec:impact}, an overview of the impact of various model components on the measured squared neutrino mass was given. The purpose is to provide a quantitative measure of their relative importance, indicating components that are negligible in the neutrino mass analysis. Among the most important effects are the radial dependencies of analyzing magnetic field and retarding potential, energy loss of signal electrons due to cyclotron motion and the Doppler broadening of the electron \tbd spectrum due to the source gas thermal motion.

The calculations presented here are implemented as part of a common C++ simulation and analysis software framework called {\sc Kasper}, which is used by the KATRIN collaboration to investigate the effect of model corrections and possible systematics, and to optimize the operational parameters of the setup for the neutrino mass measurement~\cite{PhDKleesiek2014,PhDCorona2014,PhDGroh2015,PhDFurse2015,PhDBehrens2016}.

During the ongoing commissioning measurement campaign of the KATRIN experiment, many aspects of the current response model will be verified with experimental data.
The results of recent investigations are described in \cite{Erhard2018}, \cite{Arenz2018} and \cite{Arenz2018b}.
This thorough characterization of the complex setup will allow a quantitative evaluation of the systematic effects in the neutrino mass analysis at KATRIN.

\section*{Acknowledgments}

We acknowledge the support of Helmholtz Association (HGF), Ministry for Education and Research BMBF (05A17PM3, 05A17VK2 and 05A17WO3), Helmholtz Alliance for Astroparticle Physics (HAP), and Helmholtz Young Investigator Group (VH-NG-1055) in Germany; and the Department of Energy through grants DE-AC02-05CH11231 and DE-SC0011091 in the United States.
We thank T.~Lasserre, V.~Sibille, N.~Trost and D.~V\'{e}nos for contributive discussions, and D.~Parno for her careful review and suggestions.

\appendix
\renewcommand*{\thesection}{\Alph{section}}

\section{Appendix}

\begin{table*}[tb]
    \centering
    \begin{tabular}{l r @{$\;=\;$} l}
    \toprule[\heavyrulewidth]
    \bfseries Parameter & \multicolumn{2}{l}{\hspace{3em} \bfseries Value} \\
    \midrule[\heavyrulewidth]
    Column density & $\Ncal$ & $\SI{5e17}{\per\centi\metre\squared}$ \\
    \midrule
    Active source cross-section & $A_\text{S}$ & $\SI{53}{\centi\metre\squared}$ \\
    \midrule
    Magnetic field strength (source) & $B_\text{S}$ & $\SI{3.6}{\tesla}$ \\
    Magnetic field strength (analyzing plane)  & $B_\text{A}$ & $\SI{3e-4}{\tesla}$ \\
    Magnetic field strength (maximum) & $B_\text{max}$ & $\SI{6.0}{\tesla}$ \\
    \midrule
    Inelastic scattering cross section & $\sigma_\text{inel}$ & $ \SI{3.45e-18}{\centi\meter\squared}$ \\
    \midrule
    Scattering probabilities & $P_0$ & $\SI{41.33}{\percent}$ \\
    & $P_1$ & $\SI{29.27}{\percent}$ \\
    & $P_2$ & $\SI{16.73}{\percent}$ \\
    & $P_3$ & $\hspace{0.5em} \SI{7.91}{\percent}$ \\
    & $P_4$ & $\hspace{0.5em} \SI{3.18}{\percent}$ \\
    \midrule
    Detector efficiency & $\epsilon_\text{det}$ & $0.9$ \\
    \bottomrule[\heavyrulewidth]
    \end{tabular}
    \caption[KATRIN design operational parameters]{%
    Key operational and derived parameters of KATRIN as defined in the technical design report~\cite{KATRIN2005}.
    }
    \label{tab:design}
\end{table*}

\subsection{Theoretical corrections to the \tb spectrum shape}
\label{sec:appendix:corr}

The calculation of many theoretical corrections follows the comprehensive summary of~\cite{Wilkinson1991}. Consequently, a similar nomenclature was chosen in this article.

\subsubsection{Nomenclature}

Natural units ($\hbar = c = 1$) are used unless stated otherwise.
\begin{align*}
    W &= (E + \me) / \me \\
        &\qquad \text{Electron total energy in units of } \me \\
    W_0 &= (E_0 - V_f + \me) / \me \\
        &\qquad \text{Endpoint energy in units of } \me\\
        &\qquad \text{with $V_f$ the rovibrational final state energy} \\
    p &= \sqrt{W^2-1} \\
        &\qquad \text{Electron momentum in units of } \me\\
    \beta &= p/W\\
    \alpha &= \upe^2 / \hbar c \\
        &\qquad \text{Fine structure constant} \\
    \eta &= \alpha Z / \beta, \\
        &\qquad \text{Sommerfeld parameter} \\
    \gamma &= \sqrt{1-(\alpha Z)^2} \\
    R_\text{n} &= \num{2.8840E-3} \cdot \me \\
        &\qquad \text{Nuclear radius of \chem{^3He} in units of } \me \\
    M &= \num{5497.885} \cdot \me \\
        &\qquad \text{Mass of \chem{^3He} in units of } \me \\
    \lambda_t &= |g_\text{A} / g_\text{V}| = \num{1.265} \\
        &\qquad \text{Ratio between vector and axial coupling constants}
\end{align*}
The nuclear radius $R_\text{n}$ of \chem{^3He} is given by the Elton formula~\cite{bib:elton}. The value for $\lambda_t$ is derived from the half-life of tritium by~\cite{Simkovic2008}.

\subsubsection{Fermi function}
\label{sec:appendix:fermi}

A fully relativistic description of the Fermi function is given by
\begin{equation}
    \label{eq:fermi_rel}
    F_\text{rel}(Z, W) = \frac{4}{(2 p R_\text{n})^{2(1-\gamma)}} \cdot \frac{\left| \Gamma(\gamma + \text{i}\eta) \right|^2}{\left\{ \Gamma(2\gamma\!+\!1) \right\}^2} \cdot \upe^{\pi  \eta} \; ,
\end{equation}
with the complex Gamma function $\Gamma$.
A commonly used approximate, yet sufficiently accurate for our purpose, expression for \cref{eq:fermi_rel} is~\cite{Simpson1981}
\begin{align}
    F_\text{app}(Z, W) &= F(Z, W) \cdot (\num{1.002037} - \num{0.001427} \cdot p/W)
\end{align}
with $F(Z, W)$ denoting the classical Fermi function (\cref{eq:fermi}).

\subsubsection{Radiative corrections due to virtual and real photons}

Radiative corrections, denoted by the multiplicative factor $G$, are implemented according to equation 20 of~\cite{Repko1983}:
\begin{align}
        \nonumber
    G(W, W_0) &= \biggl(W_0 - W\biggr)^{\displaystyle \frac{2\alpha}{\pi} t(\beta)} \, \\
        \nonumber
      &\quad \cdot \biggl(1 + \frac{2\alpha}{\pi} \cdot \biggl\{\, t(\beta) \; \biggl[\, \ln(2) - \frac{3}{2} + \frac{W_0-W}{W} \biggr] \\
        \nonumber
      &\quad + \frac{1}{4} \biggl[\, t(\beta) + 1 \,\biggr] \cdot \biggl[\, 2 (1+\beta^2) + 2 \ln(1 - \beta) \biggr. \\
      &\quad + \biggl. \frac{(W_0 - W)^2}{6W^2} \,\biggr] - 2 + \frac{1}{2} \beta - \frac{17}{36} \beta^2 + \frac{5}{6} \beta^3 \,\biggr\} \biggr)
\end{align}
with
\begin{equation*}
	t(\beta) = \frac{1}{\beta} \cdot \tanh^{-1}\beta -1 \; .
\end{equation*}

\subsubsection{Screening by the Coulomb field of the daughter nucleus}

The calculation of the screening correction factor $S$ follows~\cite{bib:behrens}:
\begin{equation}
    S(Z, W) = \frac{\bar{W}}{W} \left(\frac{\bar{p}}{p}\right)^{-1+2\gamma} \cdot \, \upe^{\displaystyle \pi(\bar{\eta}-\eta)} \frac{\left| \Gamma(\gamma + \text{i}\bar{\eta}) \right|^2}{\left| \Gamma(\gamma + \text{i}\eta) \right|^2} \; ,
\end{equation}
where
\begin{align*}
    \bar{W} &= W - V_0/\me \; ,\\
    \bar{p} &= \sqrt{\bar{W}^2-1} \; , \\
    \bar{\eta} &= \alpha Z \bar{W} / \bar{p} \; ,
\end{align*}
with the nuclear screening potential $V_0 = \SI{76(10)}{eV}$ of the final-state orbital electron cloud of the daughter \chem{^3He} atom after \tbd decay, as determined by~\cite{bib:hargrove}.

\subsubsection{Exchange with the orbital 1s electron}

The effect of an orbital electron exchange $I$ is calculated according to~\cite{bib:haxton}. Considering only the ground state of the daughter \chem{^3He^+} ion:
\begin{equation}
    I(Z, W) = 1 + \frac{729}{256}\,a(\tau)^2 + \frac{27}{16}\,a(\tau) \; ,
\end{equation}
where
\begin{equation*}
    a(\tau) = \exp\biggl( 2\tau \cdot \arctan\left(-\frac{2}{\tau}\right) \biggr) \left(\frac{\tau^2}{1 + \frac{1}{4} \tau^2}\right)^2 \; ,
\end{equation*}
with $\tau = -2 \alpha / p$.

\subsubsection{Recoil effects}

In the relativistic description of the three-body phase space, the spectral change due to recoil effects, including those from weak magnetism and \va interference, is reflected by the correction factor $R$~\cite{bib:bilenkii}:
\begin{equation}
    R(W,W_0) = 1 + \left(  AW-\frac{B}{W} \right) / \, C \; ,
\end{equation}
where
\begin{align*}
    A &= 2 (5\lambda_t^2 + \lambda_t \mu +1) / M \; ,\\
    B &= 2\lambda_t (\mu + \lambda_t) / M \; ,\\
    C &= 1 + 3\lambda_t^2 -b W_0 \; ,
\end{align*}
with $\mu = \num{5.107}$ being the difference between the magnetic moments of helion and triton.

\subsubsection{Finite nuclear extension}

The two correction factors $L$ and $C$, considering the finite structure of the daughter nucleus, are given by~\cite{bib:wilkinson90}. $L$ accounts for the scaling of the Coulomb field within the nucleus:
\begin{align}
        \nonumber
    L(Z,W) &= 1 + \frac{13}{60}(\alpha Z)^2 \\
        \nonumber
      &\quad - \frac{W R_\text{n}\alpha Z}{15} \cdot \frac{41-26\gamma}{2\gamma-1} \\
      &\quad - \frac{\alpha ZR\gamma}{30W} \cdot \frac{17-2\gamma}{2\gamma-1} \; .
\end{align}
The convolution of the electron and neutrino wave functions with the nucleonic wave function throughout the nuclear volume leads to $C$:
\begin{equation}
    C(Z,W) = 1 + C_0 + C_1 \cdot W + C_2 \cdot W^2 \; ,
\end{equation}
with
\begin{align*}
    C_0 &= -\frac{233}{630}(\alpha Z)^2 - \frac{1}{5}(W_0 R_\text{n})^2 + \frac{2}{35}(W_0 R_\text{n} \alpha Z) \; , \\
    C_1 &= -\frac{21}{35} R_\text{n}\alpha Z + \frac{4}{9} W_0 R_\text{n}^2 \; , \\
    C_2 &= -\frac{4}{9} R_\text{n}^2\; .
\end{align*}

\subsubsection{Recoiling Coulomb field}

The correction factor $Q$, describing the recoil of the charge distribution by the emitted lepton, is calculated according to~\cite{bib:wilkinson82}:
\begin{equation}
    Q(Z,W,W_0) = 1- \frac{\pi \alpha Z}{M p} \left( 1+ \frac{1-\lambda_t^2}{1+3\lambda_t^2} \cdot \frac{W_0-W}{3W} \right) \; .
\end{equation}


\bibliography{bibliography,references/RefArXivPublications,references/RefGeneralPublications,references/RefKATRINDiplomaMasterTheses,references/RefKATRINPhDTheses,references/RefKATRINPublications}

\end{document}